\documentstyle[epsfig,12pt]{article}
\newcommand{\be}{\begin{eqnarray}}
\newcommand{\ee}{\end{eqnarray}}

\def\lsim{\mathrel{\rlap{\lower4pt\hbox{\hskip1pt$\sim$}}
    \raise1pt\hbox{$<$}}}               
\def\gsim{\mathrel{\rlap{\lower4pt\hbox{\hskip1pt$\sim$}}
    \raise1pt\hbox{$>$}}}               

\begin{document}

\rightline{\rightline{\Large{Preprint RM3-TH/02-5}}}

\vspace{1cm}

\begin{center}

\LARGE{Comparison among Hamiltonian light-front formalisms at $q^+ = 0$ and $q^+ \neq 0$: space-like elastic form factors of pseudoscalar and vector mesons\footnote{\bf To appear in Physical Review C.}}\\

\vspace{1cm}

\large{Silvano Simula}\\

\vspace{0.5cm}

\normalsize{Istituto Nazionale di Fisica Nucleare, Sezione Roma III\\ Via della Vasca Navale 84, I-00146 Roma, Italy}

\end{center}

\vspace{1cm}

\begin{abstract}

\noindent The electromagnetic elastic form factors of pseudoscalar and vector mesons are analyzed for space-like momentum transfers in terms of relativistic quark models based on the Hamiltonian light-front formalism elaborated in different reference frames ($q^+ = 0$ and $q^+ \neq 0$). As far as the one-body approximation for the electromagnetic current operator is concerned, it is shown that the predictions of the light-front approach at $q^+ = 0$ should be preferred, particularly in case of light hadrons, because of: ~ i) the relevant role played by the $Z$-graph at $q^+ \neq 0$, and ~ ii) the appropriate elimination of spurious effects, related to the orientation of the null hyperplane where the light-front wave function is defined.

\end{abstract}

\vspace{1cm}


\vspace{0.5cm}


\newpage

\pagestyle{plain}

\section{Introduction}

\indent The Hamiltonian light-front ($LF$) formalism is one of the most popular techniques developed to treat relativistic bound states of systems containing a fixed number of constituents \cite{LF}. The $LF$ formalism is characterized by the fact that it maximizes the dimension of the subgroup of kinematical (interaction free) generators of the Poincar\`e group. As a matter of facts, the interaction term appears only in three out of the ten Poincar\`e generators, namely in the "minus" component of the four-momentum ($P^- \equiv P^0 - P^z$) and in the two transverse rotations about the $x$ and $y$ axes, with the null-plane being defined by $x^+ = t + z = 0$.

\indent Hamiltonian $LF$ quantum models have been widely used for the investigation of hadronic form factors, like, e.g., the electroweak form factors of mesons and baryons within the framework of the constituent quark picture of hadrons, or the elastic electromagnetic ($e.m.$) form factors of the deuteron viewed as a composite two-nucleon system. A relevant issue is to know to what extent the response of a composite system can be understood in terms of the properties of its constituents. To this end the one-body approximation for the current operator has been extensively considered. In case of spin-$1/2$ constituents and for the $e.m.$ current operator, which are of interest in this work, one has
 \be 
      J^{\mu} \simeq J_{(1)}^{\mu} = \sum_j \left[ f_1^{(j)}(q^2) 
      \gamma^{\mu} + f_2^{(j)}(q^2) {i \sigma^{\mu \nu} q_{\nu} \over 2 m_j} 
      \right] ~ ,
     \label{eq:one-body}
 \ee
where $q^2 = q \cdot q$ is the squared four-momentum transferred to the system and $f_{1(2)}^{(j)}$ is the Dirac(Pauli) form factors of the $j$-th constituent with mass $m_j$. In this paper we will limit ourselves to space-like $q$ (i.e., $q^2 < 0$).

\indent While the full current $J^{\mu}$ is covariant with respect to the (interaction dependent) transverse rotations, its one-body approximation (\ref{eq:one-body}) is not, and the most direct manifestation of the loss of the rotational covariance is the so-called angular condition. Indeed, it is well known \cite{LF} that all the form factors appearing in the covariant decomposition of a conserved current can be expressed in terms of the matrix elements of only one component of the current, namely the {\em plus} component $J^+ \equiv J^0 + J^z$. It may occur however that the number of form factors is less than the number of independent matrix elements of the {\em plus} component obtained from general properties of the current operator. This means that in such situations a relation among the matrix elements (the angular condition) should occur in order to constrain further their number. The use of the one-body current (\ref{eq:one-body}) may lead to important violations of the angular condition, which do not allow to extract uniquely the form factors from the matrix elements of $J_{(1)}^+$ (cf. Refs. \cite{CAR_rho,Karmanov_rho} for the case of the $\rho$-meson).

\indent Within the $LF$ formalism two different approaches \cite{DS,LPS} have been proposed to overcome the angular condition problem. Both approaches really solve the angular condition problem and make use of the {\em plus} and {\em transverse} components of the one-body current (\ref{eq:one-body}). However, the approach of Ref. \cite{DS} is realized at $q^+ = 0$ and does not introduce explicitly any covariant current, whereas the approach of Ref. \cite{LPS} is developed at $q^+ \neq 0$ and a covariant approximation of the current $J^{\mu}$ is explicitly constructed. The two approaches, which we stress are both based on the one-body approximation (\ref{eq:one-body}), are inequivalent, because the Lorentz transformation connecting a frame where $q^+ = 0$ to a frame where $q^+ \neq 0$ is interaction dependent. In other words, the impact of the additional many-body currents needed to construct the full current $J^{\mu}$, might be substantially different at $q^+ = 0$ and at $q^+ \neq 0$.

\indent The aim of this paper is to address the issue of the relevance of the many-body currents by comparing the predictions of the approaches of Refs. \cite{DS} and \cite{LPS} in case of the (space-like) $e.m.$ elastic form factors of both light and heavy pseudoscalar and vector mesons adopting the general framework of the constituent quark model. It will be shown that the two above-mentioned $LF$ approaches are inequivalent because of the different contribution of the so-called $Z$-graph \cite{Zgraph} at $q^+ = 0$ and at $q^+ \neq 0$. While at $q^+ = 0$ it is possible to cancel out exactly the $Z$-graph (see Ref. \cite{DS}), the latter is active at $q^+ \neq 0$, but ignored in Ref. \cite{LPS}. Moreover, it will be shown that the $Z$-graph provides an important contribution in case of light hadrons, whereas it vanishes in the heavy-quark limit, where the two $LF$ approaches predict the same universal Isgur-Wise ($IW$) function \cite{IW,HQS}. It will be pointed out that within the approach of Ref. \cite{LPS} the form factors are function of $Q^2 / M^2$, where $M$ is the mass of the hadron, and such a dependence is not efficient for describing the phenomenology of light hadrons. Furthermore, in case of vector mesons, the spurious effects related to the orientation of the null hyperplane where the $LF$ wave function is defined, can be properly eliminated in the $LF$ approach at $q^+ = 0$ \cite{DS}, while they are ignored and cannot be eliminated within the $LF$ approach at $q^+ \neq 0$ \cite{LPS}. Thus, the latter approach appears to maximize the impact of the additional many-body currents needed to achieve consistency with experiment, and therefore, as far as the one-body current (\ref{eq:one-body}) is concerned, the predictions of the $LF$ approach at $q^+ = 0$ should be preferred, particularly in case of light hadrons.

\indent The plan of the paper is simply as follows. Section 2 is devoted to a brief description of the two $LF$ approaches, which are then applied to the evaluation of the elastic form factor of both light and heavy pseudoscalar mesons. The case of vector mesons, where the angular condition becomes manifest, is illustrated in Section 3. The conclusions are summarized in Section 4.

\section{Pseudoscalar mesons}

\indent In this Section we consider a pseudoscalar ($PS$) meson with mass $M_{PS}$, made of two constituent quarks $q_1$ and $\bar{q}_2$ with mass $m_1$ and $m_2$, and with electric charges $e_1$ and $\bar{e}_2$. The matrix elements of the $e.m.$ current for the elastic channel are given by
 \be
       \langle P' | J^{\mu} | P \rangle = F_{PS}(Q^2) ~ (P + P')^{\mu} ~ ,
       \label{eq:FFPS} 
 \ee
where $Q^2 \equiv - q^2 = - (P' - P)^2$ is the squared four-momentum transfer ($Q^2 \geq 0$) and $F_{PS}(Q^2)$ is the elastic form factor. For both the approaches of Refs. \cite{DS} and \cite{LPS} we use the same $LF$ wave function, which as well known \cite{LF} can be factorized into the product of a center-of-mass and an intrinsic parts. In terms of the intrinsic $LF$ variables, defined as
 \be
       \xi & = & p_1^+ / P^+ = 1 - p_2^+ / P^+ ~ , \nonumber \\
       \vec{k}_{\perp} & = & \vec{p}_{1 \perp} - \xi \vec{P}_{\perp} =  
       - \vec{p}_{2 \perp} + (1 - \xi) \vec{P}_{\perp} ~ ,
       \label{eq:LFvariables}
 \ee
one gets \cite{pion,CAR_PS}
 \be
       | P \rangle_{LF} = R^{(PS)}(\xi, \vec{k}_{\perp}) ~ w_{PS}(k) ~ 
       \sqrt{ {A(\xi, \vec{k}_{\perp}) \over 4 \pi}} ~ | \vec{P}_{\perp}, 
        P^+ \rangle ~ ,
       \label{eq:wfPS}
 \ee
where $| \vec{P}_{\perp}, P^+ \rangle$ describes the $LF$ center-of-mass state [normalized according to $\langle \vec{P'}_{\perp}, P'^+ | \vec{P}_{\perp},$ $P^+ \rangle = 2P^+ ~ \delta(\vec{P'}_{\perp} - \vec{P}_{\perp}) ~ \delta(P'^+ - P^+)$], $A(\xi, \vec{k}_{\perp}) \equiv M_0 [1 - (m_1^2 - m_2^2)^2 / M_0^4] / 4 \xi (1 - \xi)$ is a normalization factor for the intrinsic $LF$ wave function, $w_{PS}(k)$ is the radial wave function (normalized as $\int_0^{\infty} dk ~ k^2 w_{PS}^2(k) = 1$) with $k \equiv \sqrt{k_{\perp}^2 + k_z^2}$, $k_z = M_0 (\xi - 1/2) + (m_2^2 - m_1^2) / 2 M_0$, and $M_0$ is the free mass given explicitly by
 \be
       M_0^2 = { m_1^2 + k_{\perp}^2 \over \xi} + { m_2^2 + k_{\perp}^2 
       \over 1 - \xi} ~ .
       \label{eq:M0}
 \ee
Finally, the quantity $R^{(PS)}$ appearing in Eq. (\ref{eq:wfPS}) is the product of (generalized) Melosh rotation spin matrices \cite{Melosh}, viz.
 \be
      \left[ R^{(PS)}(\xi, \vec{k}_{\perp}) \right]_{\lambda_1 
      \lambda_2} & = & \sum_{{\lambda'}_1 {\lambda'}_2} ~ \langle \lambda_1 
      |  R_M^{\dagger}(\xi, \vec{k}_{\perp}, m_1) | {\lambda'}_1 \rangle 
      \cdot \langle {1 \over 2} {\lambda'}_1 {1 \over 2} {\lambda'}_2 | 00 
      \rangle \nonumber \\
       & \cdot & \langle \lambda_2 | R_M^{\dagger}(1 - \xi, 
       -\vec{k}_{\perp}, m_2) | {\lambda'}_2 \rangle
       \label{eq:Melosh_00}
 \ee
with
 \be
       R_M(\xi, \vec{k}_{\perp}, m_1) = { m_1 + \xi M_0 - i \vec{\sigma} 
      \cdot (\hat{z} \times \vec{k}_{\perp} ) \over \sqrt{(m_1 + \xi M_0)^2 
      + k_{\perp}^2}}
     \label{eq:Melosh}
 \ee
In terms of the $LF$ spinors $\bar{u}(p_1, \lambda_1)$ and $v(p_2, \lambda_2)$ the Melosh factor $R^{(PS)}$ can be conveniently written as (cf. Refs. \cite{pion}(b) and \cite{DS})
 \be
       \left[ R^{(PS)}(\xi, \vec{k}_{\perp}) \right]_{\lambda_1 
       \lambda_2} = {1 \over \sqrt{2}} {1 \over \sqrt{M_0^2 - (m_1 - 
       m_2)^2}} \overline{u}(p_1, \lambda_1) \gamma^5 v(p_2,  \lambda_2) ~ .
       \label{eq:Melosh_PS}
\ee

\indent As for $w_{PS}(k)$, in what follows we will consider a specific choice, namely the eigenfunctions of the quark potential model of Ref. \cite{GI}, because the latter nicely reproduces the mass spectra of both light and heavy mesons, which are of interest in this work\footnote{Actually different choices for $w_{PS}(k)$ are clearly possible, but the qualitative results presented in this paper will not change.}. The constituent quark masses used throughout this paper are taken from Ref. \cite{GI}, namely: $m_u = m_d = 0.220 ~ GeV$, $m_s = 0.419 ~ GeV$, $m_c = 1.628 ~ GeV$ and $m_b = 4.977 ~ GeV$. The masses of the corresponding charged $PS$ and vector mesons are taken from $PDG$ \cite{PDG}, namely: $M_{\pi} ~ (M_{\rho}) = 0.1396 ~ (0.767) ~ GeV$, $M_K ~ (M_{K^*}) = 0.4937 ~ (0.892) ~ GeV$, $M_D ~ (M_{D^*}) = 1.869 ~ (2.010) ~ GeV$ and $M_B ~ (M_{B^*}) = 5.279 ~ (5.325) ~ GeV$.

\indent Let us now briefly describe the basic features of the approaches of Refs. \cite{DS} and \cite{LPS}. In the former approach the working frame is a Breit frame where $q^+ = 0$ and the $e.m.$ current operator is given by the one-body approximation (\ref{eq:one-body}).  However, the matrix elements of the one-body current do not have the decomposition given by Eq. (\ref{eq:FFPS}). This is related to the fact that the $LF$ wave function (\ref{eq:wfPS}) is specified on the null hyperplane, whose orientation can be identified by its normal four-vector $\omega$. The standard choice of $\omega$ is along the {\em minus} axis. Note that $\omega$ is a null four-vector ($\omega \cdot \omega = 0$) and $\omega \cdot q = q^+$.

\indent As firstly pointed out in Ref. \cite{Karmanov} and subsequently derived from an analysis of the Feynmann triangle diagram in Ref. \cite{DS}\footnote{The appearance of the four-vector $\omega$ in the Feynmann triangle diagram can be understood as due to the need of expressing the off-mass-shell constituent momenta, which naturally appears in any Feynmann diagram, in terms of the on-mass-shell constituent momenta, which characterize the $LF$ formalism (see Ref. \cite{DS}).}, the matrix elements of an approximate current do depend upon the four-vector $\omega$, while only the amplitudes of the full current are completely independent of $\omega$ [see Eq. (\ref{eq:FFPS})]. Then the general structure of the matrix elements of the one-body current (\ref{eq:one-body}) for an elastic process involving a $PS$ meson is given by \cite{Karmanov,DS}
 \be
       _{LF}\langle P' | J_{(1)}^{\mu} | P \rangle_{LF} = F_{PS}^{(1)}(Q^2) 
       ~ (P + P')^{\mu} + B_{PS}^{(1)}(Q^2) ~ {\omega^{\mu} \over \omega 
        \cdot P} ~ ,
       \label{eq:FF1PS} 
\ee
where $B_{PS}^{(1)}(Q^2)$ represents a spurious form factor. Since the standard choice of $\omega$ is along the {\em minus} axis, the physical form factor $F_{PS}^{(1)}(Q^2)$ can be obtained using the {\em plus} component of the one-body current (\ref{eq:one-body}). Adopting the usual Breit frame where $q^+ = q^- = 0$, one has $P'^+ = P^+ = \sqrt{M_{PS}^2 + Q^2 / 4}$ and $\vec{P'}_{\perp} = - \vec{P}_{\perp} = \vec{q}_{\perp} / 2$ with $Q^2 = q_{\perp}^2$. Assuming for sake of simplicity point-like constituents, the elastic form factor $F_{PS}^{(1)}(Q^2)$ can be cast in the following form
 \be
       F_{PS}^{(1)}(Q^2) = e_1 H_1(Q^2) + \bar{e}_2 H_2(Q^2) ~ ,
       \label{eq:F1PS}
 \ee
where \cite{DS,pion,CAR_PS}
 \be
       H_1(Q^2) & = & \int_0^1 d\xi \int d\vec{k}_{\perp} ~ \sqrt{A(\xi, 
       \vec{k}_{\perp}) ~ A(\xi, \vec{k'}_{\perp})} ~ {w_{PS}(k) ~ 
       w_{PS}(k') \over 4 \pi} \nonumber \\
       & \cdot & {\mu^2(\xi) + \vec{k}_{\perp} \cdot \vec{k'}_{\perp} \over 
       \sqrt{\mu^2(\xi) + k_{\perp}^2} \sqrt{\mu^2(\xi) + {k'}_{\perp}^2}}
       \label{eq:H_LF}
 \ee
with $\mu(\xi) \equiv m_1 ~ (1 - \xi) + m_2 ~ \xi$ and $\vec{k'}_{\perp} = \vec{k}_{\perp} + (1 - \xi) \vec{q}_{\perp}$. The explicit expression for the form factor $H_2(Q^2)$, corresponding to the coupling of the virtual photon with the antiquark $\bar{q}_2$, can be easily obtained from Eq. (\ref{eq:H_LF}) by using simply $\vec{k'}_{\perp} = \vec{k}_{\perp} - \xi \vec{q}_{\perp}$. Note that the form factor (\ref{eq:F1PS}) is a function of $Q^2$ and of the constituent masses $m_1$ and $m_2$.

\indent In case of the approach of Ref. \cite{LPS} the working frame is a special Breit frame where the four-momentum transfer $q$ is along the spin-quantization axis, i.e. the $z$ axis. Therefore one has $\omega \cdot q = q^+ \neq 0$, more precisely: $q^+ = - q^- = Q$, $\vec{q}_{\perp} = 0$, $P^+ = \sqrt{M_{PS}^2 + Q^2/4} - Q/2$, $P'^+ = \sqrt{M_{PS}^2 + Q^2/4} + Q/2$ and $\vec{P}_{\perp} = \vec{P'}_{\perp} = 0$. In such a special frame and in case of elastic processes a choice $j^{\mu}$ for the e.m. current operator compatible with (extended) Poincar\`e covariance and hermiticity is given by \cite{LPS}
 \be
       j^{\mu} = {1 \over 2} \left\{ C^{\mu} + L^{\mu \nu}[r_x(-\pi)] ~ e^{i 
       \pi S_x} ~ C_{\nu}^* ~ e^{-i \pi S_x} \right\} ~ ,
      \label{eq:LPS}
 \ee
where $r_x(-\pi)$ represents a ($-\pi$) rotation around the $x$ axis, $S_x$ is the $x$-component of the $LF$ spin operator and $L(\ell)$ is the element of the Lorentz group corresponding to $\ell ~ \varepsilon ~ SL(2,C)$. In Eq. (\ref{eq:LPS}) the operator $C^{\mu}$ should fulfill the (extended) Poincar\`e covariance and the specific choice made in Ref. \cite{LPS} is as follows:
 \be
       C^+ & = & J_{(1)}^+ ~ , \nonumber \\
       \vec{C}_{\perp} & = & \vec{J}_{(1) \perp} ~ , \nonumber \\
       C^- & = & J_{(1)}^+ ~ ,
       \label{eq:LPS_J}
 \ee
where the last equation ensures the gauge invariance of the current operator $j^{\mu}$. The covariant decomposition of $_{LF}\langle P' | j^{\mu} | P \rangle_{LF}$ is therefore given by
 \be
       _{LF}\langle P' | j^{\mu} | P \rangle_{LF} =  {\cal{F}}_{PS}^{(1)} ~ 
       (P + P')^{\mu} ~ ,
       \label{eq:FF_LPS}
 \ee
where no spurious structure is present thanks to the gauge invariance of $j^{\mu}$ and to the fact that $\omega \cdot q = q^+ \neq 0$. Then, the elastic form factor ${\cal{F}}_{PS}^{(1)}$ can be calculated using the {\em plus} component $j^+$, whose matrix element $_{LF}\langle P' | j^+ | P \rangle_{LF}$ reduces to the corresponding one of the {\em plus} component of the one-body current $J_{(1)}^+$ evaluated at $q^+ = - q^- = Q$ (see Ref. \cite{LPS}). Therefore, one gets
 \be
       {\cal{F}}_{PS}^{(1)} = e_1 {\cal{H}}_1(Q^2 / M_{PS}^2) + \bar{e}_2 
       {\cal{H}}_2(Q^2 / M_{PS}^2)
       \label{eq:F_LPS}
 \ee
with
 \be
       {\cal{H}}_1(Q^2 / M_{PS}^2) & = & { 1 - \kappa \over 1 - \kappa / 2} 
       ~ \int_0^1 d\xi \int d\vec{k}_{\perp} ~ \sqrt{A(\xi, \vec{k}_{\perp}) 
       ~ A(\xi', \vec{k}_{\perp})} \nonumber \\
       & \cdot & {w_{PS}(k) ~ w_{PS}(k') \over 4 \pi}  ~ {\mu(\xi) \mu(\xi') 
       + k_{\perp}^2 \over \sqrt{\mu^2(\xi) + k_{\perp}^2} \sqrt{\mu^2(\xi') 
       + k_{\perp}^2}} ~ ,
       \label{eq:H_LPS}
 \ee
where $\xi' = \kappa + (1 - \kappa) ~ \xi$ and
 \be
       \kappa \equiv { q^+ \over P'^+} = {Q / M_{PS} \over \sqrt{1 + Q^2 / 4 
       M_{PS}^2} + Q / 2 M_{PS}} = \sqrt{{Q^4 \over 4M_{PS}^4} + {Q^2 \over 
       M_{PS}^2}} - {Q^2 \over 2 M_{PS}^2} ~ .
      \label{eq:kappa}
 \ee 
The explicit expression for the form factor ${\cal{H}}_2(Q^2 / M_{PS}^2)$ can be easily obtained from Eq. (\ref{eq:H_LPS}) by using simply $\xi' = (1 - \kappa) ~ \xi$. Note that the factor $(1 - \kappa) / (1 - \kappa / 2)$ in the $r.h.s.$ of Eq. (\ref{eq:H_LPS}) is nothing else than the factor $2 P^+ / (P^+ + P'^+)$, where the numerator comes from the normalization of the $LF$ center-of-mass states. Moreover, one has $0 \leq \kappa \leq 1$.

\indent In Eq. (\ref{eq:F_LPS}) we have explicitly taken into account that within the approach of Ref. \cite{LPS} the elastic form factor ${\cal{F}}_{PS}^{(1)}$ is not a function of $Q^2$, but of the ratio $Q^2 / M_{PS}^2$ (and of the constituent masses $m_1$ and $m_2$). This is an important feature, which naturally emerges when a "longitudinal" frame with $q^+ \neq 0$ and $\vec{q}_{\perp} = 0$ is chosen. As a matter of facts, since only the $LF$ fraction $\xi$ changes in the final state, the form factor ${\cal{F}}_{PS}^{(1)}$ cannot be function of $q^+$ only, but of $q^+ / P^+$ (or equivalently $q^+ / P'^+ = \kappa$). Thus, the dependence of the elastic form factors upon the ratio between the momentum transfer and the mass of the system is expected to characterize the predictions of the $LF$ approach of Ref. \cite{LPS} for a generic hadron\footnote{This means that the $LF$ approach at $q^+ \neq 0$ provides naturally implicit many-body contributions associated with the hadron mass.} and this fact has important consequences. The most direct one is that the charge radius of a hadron is expected to be approximately proportional to the inverse of its mass. Such a behavior is completely at variance with the phenomenology of light hadrons. Indeed, for instance, the experimental charge radii of the pion, the kaon and the nucleon are approximately the same [$r_{ch}^{\pi} = 0.660 \pm 0.024 ~ fm$ \cite{radii}(a), $r_{ch}^K = 0.58 \pm 0.04 ~ fm$ \cite{radii}(b) and $r_{ch}^p = 0.883 \pm 0.014 ~ fm$ \cite{radii}(c)], while the kaon (nucleon) mass is larger than the pion mass by a factor of $\simeq 3.5$ ($6.7$).

\indent From the above considerations it is quite clear that the approach of Ref. \cite{LPS} is likely to be not efficient for the description of the phenomenology of light hadrons; in other words, substantial effects of additional many-body currents should be explicitly considered in order to achieve consistency with experiment. An explicit demonstration is provided by the pion case. The results obtained using Eqs. (\ref{eq:F_LPS}-\ref{eq:H_LPS}), are reported in Fig. 1 and compared with the corresponding results of the approach of Ref. \cite{DS} based on Eqs. (\ref{eq:F1PS}-\ref{eq:H_LF}). The form factor evaluated at $q^+ \neq 0$ exhibits a very rapid falloff with increasing $Q^2$, corresponding to a charge radius of $\simeq 5 ~ fm$. On the contrary, the form factor obtained at $q^+ = 0$ is much higher and corresponds to a charge radius of $\simeq 0.46 ~ fm$. Similar results hold as well in case of the kaon, whose charge radius turns out to be $\simeq 2 ~ fm$ at $q^+ \neq 0$ and $\simeq 0.43 ~ fm$ at $q^+ = 0$. All the calculations have been done assuming point-like constituents and therefore only within the approach at $q^+ = 0$ it is possible to recover the agreement with the experimental charge radius of the pion (and the kaon) by introducing a constituent size of $\simeq 0.45 ~ fm$, as already proposed in Ref. \cite{CAR_PS}. The same constituent size is suggested also by the recent analysis of the elastic nucleon form factors carried out within the covariant $LF$ approach at $q^+ = 0$ in Ref. \cite{SIM01}.

\begin{figure}[htb]

\vspace{0.25cm}

\centerline{\epsfxsize=12cm \epsfig{file=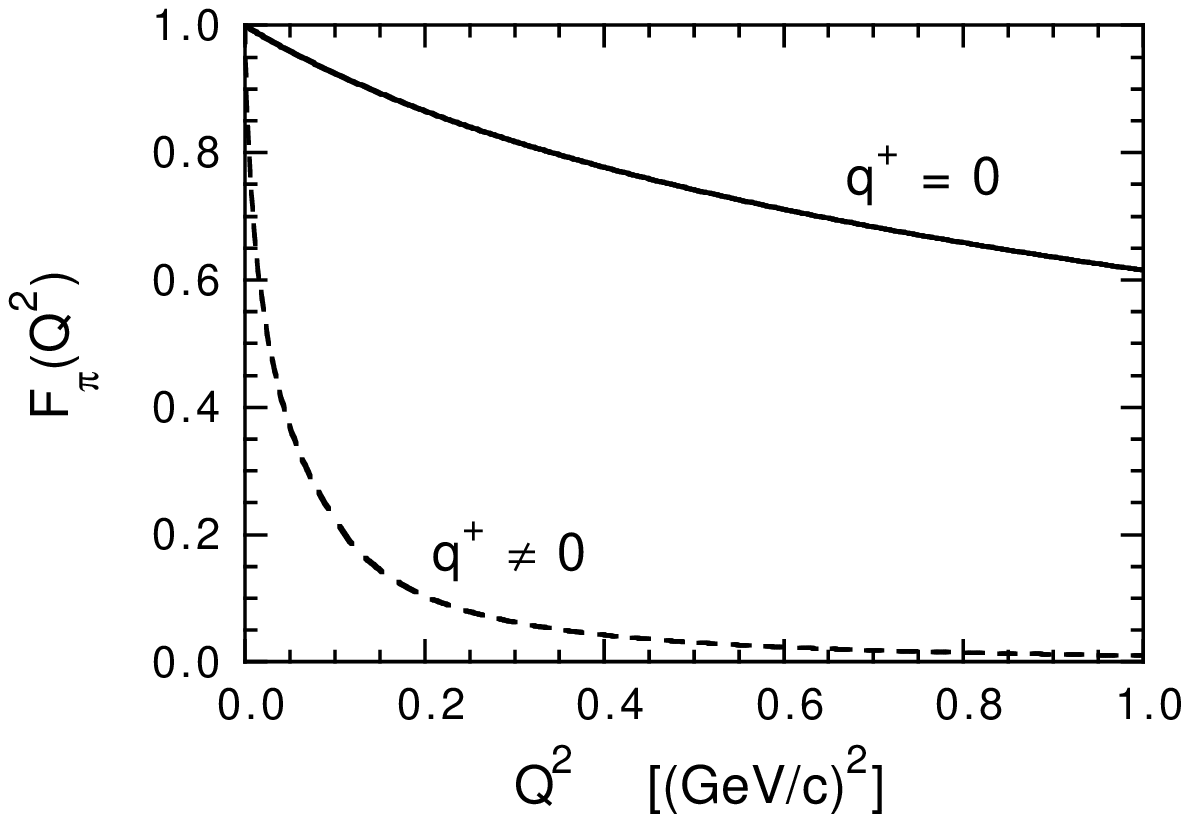}}

{\small \noindent {\bf Figure 1}. Elastic form factor of the pion, $F_{\pi}(Q^2)$, versus $Q^2$. Solid line: $LF$ approach of Ref. \protect\cite{DS} at $q^+ = 0$, corresponding to Eqs. (\protect\ref{eq:F1PS}-\protect\ref{eq:H_LF}). Dashed line: $LF$ approach of Ref. \protect\cite{LPS} at $q^+ \neq 0$, corresponding to Eqs. (\protect\ref{eq:F_LPS}-\protect\ref{eq:H_LPS}). Point-like constituent quarks are assumed in the calculations. The radial wave function $w_{PS}(k)$ is taken to be the eigenfunction of the quark potential model of Ref. \protect\cite{GI}.}

\vspace{0.25cm}

\end{figure}

\indent The question to be addressed now is clearly the origin of the large differences in the pion form factor evaluated at $q^+ = 0$ and $q^+ \neq 0$. The answer is already well known from the works of Ref. \cite{Zgraph} and more recently from the results of Refs. \cite{DS,SIM_Z}. There it has been shown that the one-body $LF$ form factor at $q^+ = 0$ [Eqs. (\ref{eq:F1PS}-\ref{eq:H_LF})] matches exactly the form factor embedded in the Feynmann triangle diagram, evaluated using the one-body current (\ref{eq:one-body}) and adopting the appropriate bound-state vertex corresponding to the $LF$ wave function (\ref{eq:wfPS}) [see Eq. (11) of Ref. \cite{DS}]. The matching is due to the vanishing of the contribution of the $Z$-graph at $q^+ = 0$. Thus, one has $F_{PS}^{(1)} = F_{PS}^{(triangle)}$. Thanks to the covariance property of the triangle diagram, the same form factor can be obtained in a frame where $q^+ \neq 0$. However, in such a frame the anatomy of the form factor is different in the sense that it is given by the sum of two non-vanishing contributions, the spectator and the $Z$-graph terms: $F_{PS}^{(triangle)} = {\cal{F}}_{PS}^{(sp.)} +  {\cal{F}}_{PS}^{(Z-graph)}$. The spectator term (see Ref. \cite{SIM_Z}), evaluated in the Breit frame where $q^+ = - q^- = Q$, turns out to coincide with Eqs. (\ref{eq:F_LPS}-\ref{eq:H_LPS}), i.e. with the form factor predicted by the approach of Ref. \cite{LPS}: ${\cal{F}}_{PS}^{(1)} = {\cal{F}}_{PS}^{(sp.)}$. Thus, finally one has
 \be
        F_{PS}^{(1)} = {\cal{F}}_{PS}^{(1)} + {\cal{F}}_{PS}^{(Z-graph)} ~ ,
        \label{eq:Zgraph}
 \ee
where ${\cal{F}}_{PS}^{(Z-graph)}$ is the contribution of the $Z$-graph in the triangle diagram evaluated in the Breit frame where $q^+ = - q^- = Q$. Thus, the origin of the differences in the pion form factor evaluated within the $LF$ approaches of Refs. \cite{DS} and \cite{LPS} is the $Z$-graph contribution at $q^+ \neq 0$, which is ignored in Ref. \cite{LPS}. From Fig. 1 it is clear that the $Z$-graph dominates the pion form factor at $q^+ \neq 0$, while the contribution of the spectator term is almost negligible (except at the photon point). Therefore, since the $Z$-graph is a many-body process, we can conclude that the approach of Ref. \cite{LPS}, based on the specific choice (\ref{eq:LPS_J}) and on the Breit frame where $q^+ = - q^- = Q$, appears {\em to maximize the impact of the many-body currents needed for consistency with experiment}, particularly in case of light hadrons.

\indent Since the $Z$-graph is expected to vanish in the heavy-quark limit, it is worthwhile to study the behavior of the two $LF$ approaches in that limit. We expect that for fixed values of the dot product $w$ of the initial and final meson four-velocities, given by
 \be
       w = P \cdot P' / M_{PS}^2 = 1 + Q^2 / 2 M_{PS}^2 ~, 
       \label{eq:w}
 \ee
one should have
 \be
       \mbox{lim}_{m_1 \to \infty} ~ H_1(Q^2) = \mbox{lim}_{m_1 \to \infty} 
       ~ {\cal{H}}_1(Q^2 / M_{PS}^2) = \xi_{IW}(w) ~ ,
       \label{eq:HQlimit_PS}
 \ee
where $\xi_{IW}(w)$ is the Isgur-Wise ($IW$) form factor \cite{IW,HQS}, which has been already calculated within the $LF$ approach at $q^+ = 0$ in Ref. \cite{SIM_IW}. We have therefore calculated the form factors $H_1(Q^2)$ [Eq. (\ref{eq:H_LF})] and ${\cal{H}}_1(Q^2 / M_{PS}^2)$ [Eq. (\ref{eq:H_LPS})] for various values of the constituent mass $m_1$ pertaining to the cases of $\pi$, $K$, $D$ and $B$ mesons, keeping fixed the constituent mass $m_2$ at the value $m_2 = 0.220 ~ GeV$. The results are reported in Fig. 2 in terms of the variable $w$ and compared with the $IW$ function $\xi_{IW}(w)$ calculated in Ref. \cite{SIM_IW}.

\begin{figure}[htb]

\vspace{0.25cm}

\centerline{\epsfxsize=16cm \epsfig{file=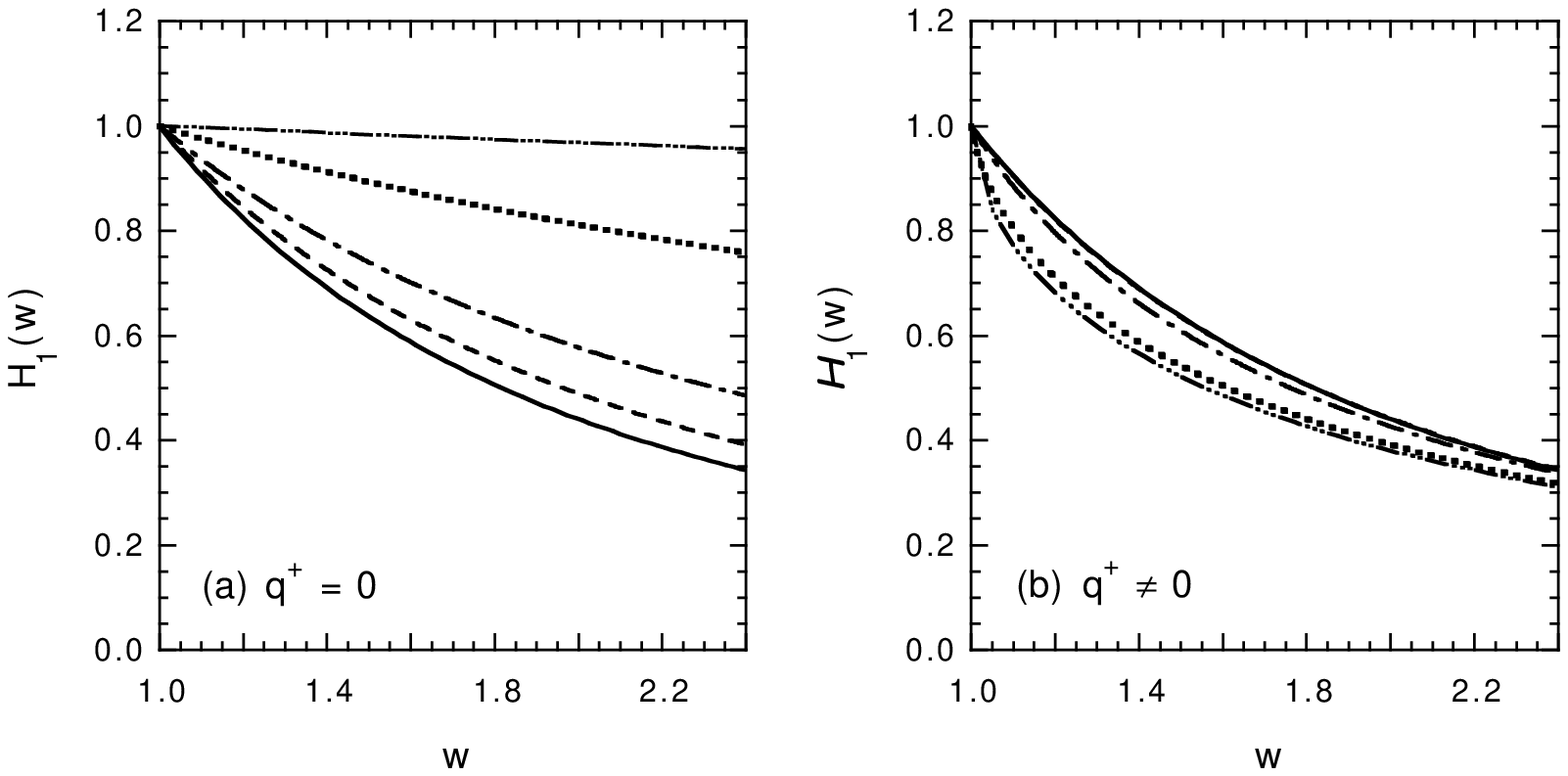}}

{\small \noindent {\bf Figure 2}. The $PS$ form factors $H_1$ [Eq. (\protect\ref{eq:H_LF})] (a) and ${\cal{H}}_1$ [Eq. (\protect\ref{eq:H_LPS})]  (b) versus the variable $w$ [Eq. (\protect\ref{eq:w})]. Triple-dot-dashed, dotted, dot-dashed and dashed lines correspond to the cases of $\pi$, $K$, $D$ and $B$ mesons, respectively. In (a) and (b) the solid line is the same $IW$ function $\xi_{IW}(w)$ as calculated in Ref. \protect\cite{SIM_IW}. In (b) the dashed line almost coincides with the solid line. Point-like constituent quarks are assumed in the calculations. The radial wave functions $w_{PS}(k)$ are taken to be the eigenfunctions of the quark potential model of Ref. \protect\cite{GI}.}

\vspace{0.25cm}

\end{figure}

\indent Few comments are in order: ~ i) Eq. (\ref{eq:HQlimit_PS}) is fulfilled, i.e. the two $LF$ approaches predict the same asymptotic $IW$ function. This confirms that the differences between the results at $q^+ = 0$ and $q^+ \neq 0$ are entirely due to the $Z$-graph contribution ignored at $q^+ \neq 0$; ~ ii) the convergence of the calculated form factors to the asymptotic $IW$ function occurs from below at $q^+ \neq 0$ and from above at $q^+ = 0$ and it is much faster in the former case: ${\cal{H}}_1 \simeq \xi_{IW}$ already at the B-meson mass (see Fig. 2(b)); ~ iii) the dependence of the calculated form factor at $q^+ \neq 0$ upon the constituent mass $m_1$ is quite mild at $q^+ \neq 0$, which implies that in terms of the variable $w$ the approach of Ref. \cite{LPS} predicts quite similar form factors for light and heavy mesons. Such a prediction is not reasonable; indeed, the dynamics of a heavy quark is characterized by the Heavy Quark Symmetry ($HQS$) \cite{IW,HQS}, which makes a heavy quark blind to the spin and flavor of light spectator quarks. The situation is opposite in light hadrons, where already the mass spectrum clearly exhibits a rich spin and flavor dependent structure. Thus, at $q^+ \neq 0$ the form factors of light hadrons are basically dominated the $Z$-graph, while the one-body spectator term [i.e., Eqs. \ref{eq:F_LPS}-\ref{eq:H_LPS})] plays only a marginal role. On the contrary, at $q^+ = 0$ the $Z$-graph is suppressed and the main contribution to the form factors is given by the one-body spectator term [i.e., Eqs. \ref{eq:F1PS}-\ref{eq:H_LF})]. Therefore, the approach of Ref. \cite{LPS} should be improved by making a choice of the operator $C^{\mu}$ different from Eq. (\ref{eq:LPS_J}). If this will be done in such a way to include the effects of the $Z$-graph, we expect that the present one-body predictions at $q^+ = 0$ will be recovered as the sum of one-body and additional many-body contributions at $q^+ \neq 0$.

\indent If we scale the result obtained for $r_{ch}^{\pi}$ at $q^+ \neq 0$ (i.e. $r_{ch}^{\pi} \sim 5 ~ fm$) according to the $1 / M$ dependence, where $M$ is the mass of the system, one may obtain a charge radius of  $\sim 0.7 ~ fm$ for $M \sim 1 ~ GeV$, i.e. around the nucleon mass. In other words it might happen that the approach of Ref. \cite{LPS} could reproduce the proton charge radius even assuming point-like (or almost point-like) constituents. Moreover, thanks to the approximate dipole behavior of the proton form factors at least for $Q^2 \lsim M^2 \simeq 1 ~ GeV^2$,  a good agreement with the experimental data at low $Q^2$ might be achieved at $q^+ \neq 0$ without the need of introducing a finite size for the constituent quarks. However, it should be clear that such an agreement is not physically meaningful unless one demonstrates that the $Z$-graph gives a negligible contribution. This is likely not to be the case as shown in Fig. 3, where the quantity $R_Z$, defined as
 \be
       R_Z = {H_1 - {\cal{H}}_1 \over {\cal{H}}_1} ~ ,
       \label{eq:RZ}
 \ee
is reported as a function of the variable $w$ for the various pseudoscalar mesons so far considered. The quantity $R_Z$ is a measure of the relative importance of the $Z$-graph with respect to the calculated form factor at $q^+ \neq 0$. Since $M_K$ is almost half of the nucleon mass and $M_D$ approximately twice the nucleon mass, we expect that, as a conservative estimate, the $Z$-graph contribution for a system with a mass around $1 ~ GeV$ should be between the dotted and dashed lines shown in Fig. 3, and therefore not negligible. 

\begin{figure}[htb]

\vspace{0.25cm}

\centerline{\epsfxsize=12cm \epsfig{file=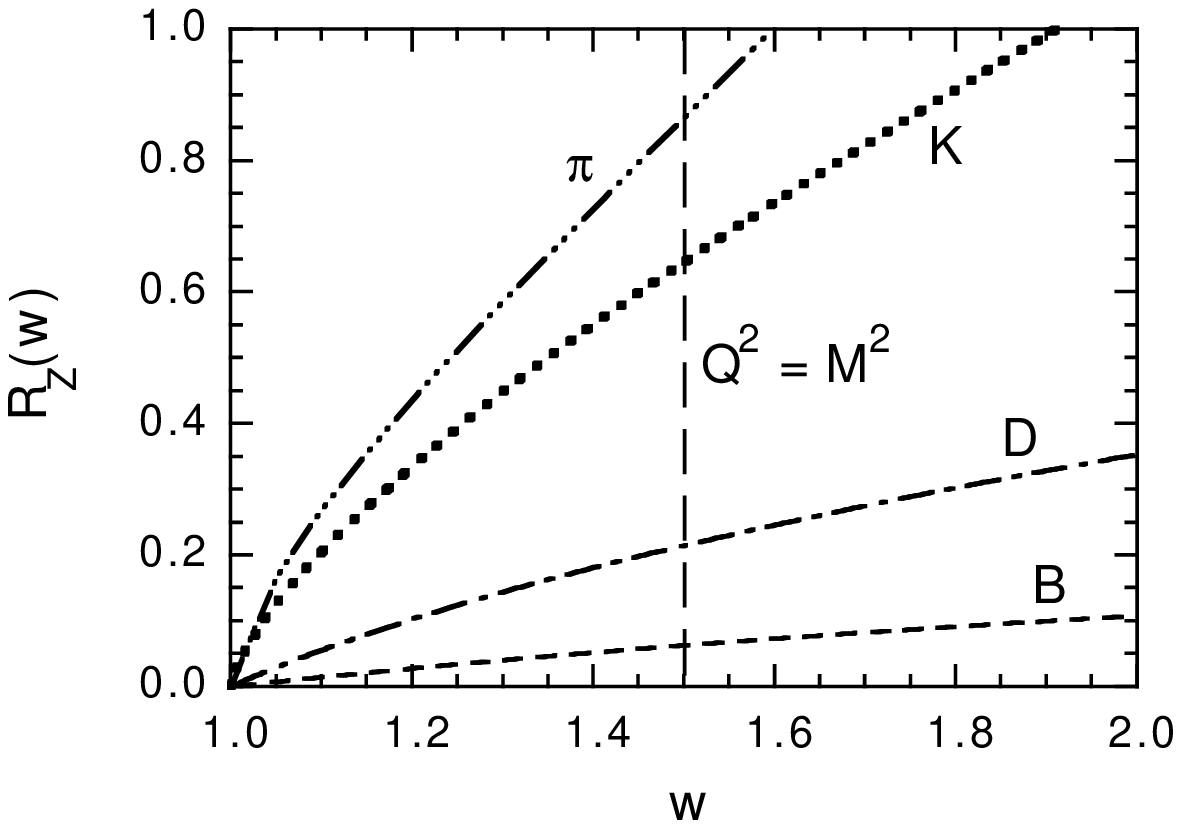}}

{\small \noindent {\bf Figure 3}. The ratio $R_Z$ [Eq. (\protect\ref{eq:RZ})] obtained using the results shown in Fig. 2, versus the variable $w$ [Eq. (\protect\ref{eq:w})]. The meaning of the lines is the same as in Fig. 2. The vertical long-dashed line at $w = 1.5$ corresponds to $Q^2 = M^2$, where $M$ is the mass of the system.}

\vspace{0.25cm}

\end{figure}

\indent The above criticisms can be directly extended to the recent results obtained in Ref. \cite{PFSA} for the nucleon elastic form factors within the so-called Point Form Spectator Approximation ($PFSA$). Such an approach was firstly proposed and illustrated in Ref. \cite{Klink}; here it suffices to say that it is based on the one-body approximation (\ref{eq:one-body}) for the $e.m.$ current operator, but carried out within a form of the dynamics different from the $LF$ one, namely the point form. However, as in the case of the $LF$ approach at $q^+ \neq 0$, within the $PFSA$ the elastic form factors of a hadron are not function of $Q^2$, but of $Q^2 / M^2$, where $M$ is the mass of the hadron (cf. Ref. \cite{Klink}). Moreover, within the $PFSA$ the $Z$-graph is totally ignored (i.e., no match with the Feynmann triangle diagram). Therefore, the agreement with the elastic nucleon data obtained in Ref. \cite{PFSA} assuming point-like constituent quarks, is likely to be physically meaningless in the same way and with the same arguments already explained in case of the $LF$ approach at $q^+ \neq 0$. A more detailed analysis of the $PFSA$ is in progress.

\indent Before closing this Section, we want to stress again the main result so far achieved, namely: among various Hamiltonian formalisms (point as well as light-front forms) based on the one-body $e.m.$ current (\ref{eq:one-body}), the predictions of the $LF$ approach at $q^+ = 0$  should be preferred, particularly in case of light hadrons. This is due to the fact that when $q^+ \neq 0$ the relevance of many-body currents appears to be amplified by the occurrence of the $Z$-graph term, which turns out to be essential also to compensate the unwanted $Q^2 / M^2$ dependence of the one-body form factors calculated in "longitudinal" frames. Finally, we point out that the choice $q^+ = 0$ is possible only for space-like $q$ ($q^2 \leq 0$). Indeed, for time-like $q$ ($q^2 > 0$) one has always $q^+ \neq 0$. In this case one needs to perform an analytic continuation from space-like to time-like $q$. This cannot be easily done in the standard $LF$ formalism because the contribution of the $Z$-graph cannot be eliminated when $q^2 > 0$ (see Refs. \cite{SIM_Z,Bmeson}). However, a proper analytic continuation can be achieved for the Feynmann triangle diagram  by means of the so-called dispersion approach, which is described in Ref. \cite{Dima} and has been extensively applied to time-like processes, like heavy meson weak decays, in Ref. \cite{MNS}. We stress that for space-like $q$ the dispersion approach result matches the $LF$ one at $q^+ = 0$ (see Ref. \cite{Dima}).

\section{Vector mesons}

\indent The elastic $e.m.$ response of a vector system is described by three physical form factors $F_i(Q^2)$ ($i = 1, 2, 3$), which appear in the following covariant decomposition of the matrix elements of the $e.m.$ current operator $J^{\mu}$:
\be
       J^{\mu}(s', s) & = & - (P + P')^{\mu} \left\{ F_1(Q^2) e^*(P', s')  
       \cdot e(P, s) + {F_2(Q^2) \over 2M_V^2} \left[ e^*(P', s') \cdot q 
        \right] \left[ e(P, s) \cdot q \right] \right\} 
       \nonumber \\
       & + & F_3(Q^2) \left\{ \left[ e^{\mu}(P', s') \right]^* \left[  e(P, 
       s) \cdot q \right] - e^{\mu}(P, s) \left[ e^*(P', s') \cdot q \right]  
       \right\} ~,
       \label{eq:FFV}
 \ee
where $e(P, s)$ is the $LF$ polarization four-vector of the spin-1 system (with mass $M_V$) corresponding to spin projection $s$ and total four-momentum $P$.

\indent In the standard $LF$ approach at $q^+ = 0$ (see Ref. \cite{CAR_rho}) only the matrix elements of the {\em plus} component of the one-body approximation (\ref{eq:one-body}), which will be denoted by $J_{(1)}^+(s', s)$, are taken into account. After considering general properties of the current operator (like, e.g., the time reversal symmetry) the number of independent matrix elements $J_{(1)}^+(s', s)$ turns out to be four, while the physical form factors are three. A further condition arises from the rotational invariance of the charge density, which however involves transformations based upon Poincar\`e generators depending on the interaction. Such an additional constraint, known as the angular condition \cite{GK}, reads as 
 \be
       (1 + 2 \eta) J_{(1)}^+(1, 1)+ J_{(1)}^+(1, -1) - \sqrt{8 \eta} 
       J_{(1)}^+(1, 0) - J_{(1)}^+(0, 0) = 0 ~ ,
       \label{eq:angular}
 \ee
where $\eta \equiv Q^2 / 4 M_V^2$. The angular condition (\ref{eq:angular}) is not satisfied by the matrix elements $J_{(1)}^+(s', s)$ and therefore the extraction of the one-body form factors $F_i^{(1)}(Q^2)$ is not unique.

\indent In Ref. \cite{DS} the problem of the violation of the angular condition (\ref{eq:angular}) is solved by considering the matrix elements of the tensor $T_{(1)}^{\mu, \alpha \beta}$, which are related to those of the one-body current by
 \be
       J_{(1)}^{\mu}(s', s) = e_{\alpha}^*(P', s') ~ T_{(1)}^{\mu, \alpha 
       \beta} ~ e_{\beta}(P, s) ~ .
       \label{eq:tensor}
 \ee
In terms of spin-1 $LF$ wave function one has
 \be
       T_{(1)}^{\mu, \alpha \beta} = ~ _{LF}\langle P', \alpha | 
       J_{(1)}^{\mu} | P,  \beta \rangle_{LF} ~ ,
       \label{eq:tensor_LF}
 \ee
where for an $S$-wave vector system
 \be
        | P, \beta \rangle_{LF} = \mbox{R}^{(V)}(\xi, \vec{k}_{\perp}; 
        \beta) ~ w_V(k) ~ \sqrt{{A(\xi, \vec{k}_{\perp}) \over 4\pi}} ~ | 
        \vec{P}_{\perp}, P^+ \rangle
        \label{eq:wfV}
 \ee
with
 \be
       \left[ \mbox{R}^{(V)}(\xi, \vec{k}_{\perp}; \beta) \right]_{\lambda_1 
       \lambda_2} & = & {1 \over \sqrt{2}} {1 \over  \sqrt{M_0^2 - (m_1 - 
       m_2)^2}} \nonumber \\
       & \cdot & \bar{u}(p_1, \lambda_1) \left[ \gamma^{\beta} - {(p_1 -  
       p_2)^{\beta} \over M_0 + m_1 + m_2} \right] v(p_2, \lambda_2) ~.
       \label{eq:Melosh_V}
 \ee
Following Refs. \cite{DS} and \cite{Karmanov_rho,Karmanov} the general decomposition of the tensor $T_{(1)}^{\mu, \alpha \beta}$ reads as
 \be
       T_{(1)}^{\mu, \alpha \beta} = I_{(1)}^{\mu, \alpha \beta} + 
       B_{(1)}^{\mu, \alpha \beta}(\omega) ~ ,
       \label{eq:T1}
 \ee
where the tensor $I_{(1)}^{\mu, \alpha \beta}$ is independent of the four-vector $\omega$, while $B_{(1)}^{\mu, \alpha \beta}(\omega)$ contains all the possible covariant structures depending on $\omega$. One gets \cite{Karmanov_rho,DS}
 \be
        I_{(1)}^{\mu, \alpha \beta} & = & - (P + P')^{\mu} \left\{ 
        F_1^{(1)}(Q^2) \left[ g^{\alpha \beta} - {P^{\alpha} P^{\beta} \over 
        M_V^2} - {{P'}^{\alpha} {P'}^{\beta} \over M_V^2} + {{P'}^{\alpha} 
        P^{\beta} \over M_V^2} {P \cdot P' \over M_V^2} \right] \right. 
        \nonumber \\
        & + & \left. {F_2^{(1)}(Q^2) \over 2 M_V^2} \left( q^{\alpha} - {P' 
        \cdot q \over M_V^2} {P'}^{\alpha} \right) \left( q^{\beta} - {P 
        \cdot q \over M_V^2} P^{\beta} \right) \right\} \nonumber \\
        & + & F_3^{(1)}(Q^2) \left\{ \left( g^{\mu \alpha} - {{P'}^{\mu} 
        {P'}^{\alpha} \over M_V^2} \right) \left( q^{\beta} - {P \cdot q 
        \over M_V^2} P^{\beta} \right) \right. \nonumber \\
        & - & \left. \left( g^{\mu \beta} - {P^{\mu} P^{\beta} \over M_V^2} 
        \right) \left( q^{\alpha} - {P' \cdot q \over M_V^2} {P'}^{\alpha} 
        \right) \right\} \nonumber \\
        & - & (P + P')^{\mu} \left\{ H_1^{(1)}(Q^2) {{P'}^{\alpha} P^{\beta} 
        \over M_V^2} + {H_2^{(1)}(Q^2) \over 2 M_V^2} \left(q^{\alpha} 
        P^{\beta} - q^{\beta} {P'}^{\alpha} \right) \right\} \nonumber \\
        & + & H_3^{(1)}(Q^2) \left( g^{\mu \alpha} P^{\beta} + g^{\mu \beta} 
        {P'}^{\alpha} \right) + H_4^{(1)}(Q^2) q^{\mu} {q^{\alpha} P^{\beta} 
        + q^{\beta} {P'}^{\alpha} \over M_V^2}
        \label{eq:F+H}
 \ee
and
 \be
       B_{(1)}^{\mu, \alpha \beta}(\omega) & = & {M_V^2 \over 2 (\omega 
       \cdot P)} \omega^{\mu} ~ \left[ B_1^{(1)}(Q^2) ~ g^{\alpha \beta} + 
       B_2^{(1)}(Q^2) ~ {q^{\alpha} q^{\beta} \over M_V^2} + M_V^2 
       B_3^{(1)}(Q^2) ~ {\omega^{\alpha} \omega^{\beta} \over (\omega \cdot 
       P)^2} \right. \nonumber \\
       & + & \left. B_4^{(1)}(Q^2) ~ {q^{\alpha} \omega^{\beta} - q^{\beta} 
       \omega^{\alpha} \over 2 (\omega \cdot P)} \right] + (P + P')^{\mu} 
       \left[ M_V^2 B_5^{(1)}(Q^2) ~ {\omega^{\alpha} \omega^{\beta} \over 
       (\omega \cdot P)^2} \right. \nonumber \\
       & + & \left. B_6^{(1)}(Q^2) ~ {q^{\alpha} \omega^{\beta} - q^{\beta} 
       \omega^{\alpha} \over 2 (\omega \cdot P)} \right] + {M_V^2 \over 
       (\omega \cdot P)} ~ B_7^{(1)}(Q^2) ~ \left[ \left( g^{\mu \alpha}  
       - {q^{\mu} q^{\alpha} \over q^2} \right) \omega^{\beta} \right. 
       \nonumber \\
       & + & \left. \left( g^{\mu \beta} - {q^{\mu} q^{\beta} \over q^2} 
       \right) \omega^{\alpha} \right] + B_8^{(1)}(Q^2) ~ q^{\mu} ~ 
       {q^{\alpha} \omega^{\beta} + q^{\beta} \omega^{\alpha} \over 2 
       (\omega \cdot P)} ~ ,
       \label{eq:BV_LF}
 \ee
where all the covariant structures included in Eqs. (\ref{eq:F+H}-\ref{eq:BV_LF}) satisfy both parity and time reversal symmetries.

\indent In Eq. (\ref{eq:F+H}) there are seven form factors, namely the three form factors $F_i^{(1)}(Q^2)$ ($i = 1, 2, 3$) and the four form factors $H_j^{(1)}(Q^2)$ ($j = 1, ..., 4$). The form factors $F_i^{(1)}(Q^2)$ appear in covariant structures which are transverse to all the external momenta $P$, $P'$ and $q$, while the form factors $H_j^{(1)}(Q^2)$ describe the loss of transversity (including the possible loss of gauge invariance) of the tensor (\ref{eq:F+H}). Therefore, since $e(P, s) \cdot P = e(P', s') \cdot P' = 0$, the form factors $H_j^{(1)}(Q^2)$ do not appear in the decomposition of the matrix elements $J_{(1)}^{\mu}(s', s)$.

\indent In Eq. (\ref{eq:BV_LF}) all the $B_k^{(1)}(Q^2)$ ($k = 1, 2, ... 8$) are spurious form factors, which can contribute to the matrix elements $J_{(1)}^{\mu}(s, s')$, namely:
 \be
       J_{(1)}^{\mu}(s', s) & = & - (P + P')^{\mu} \left\{ F_1^{(1)}(Q^2) 
       e^*(P', s') \cdot e(P, s) + {F_2^{(1)}(Q^2) \over 2M_V^2} \left[ 
       e^*(P', s') \cdot q \right] \left[ e(P, s) \cdot q \right] \right\} 
       \nonumber \\
       & + & F_3^{(1)}(Q^2) \left\{ \left[ e^{\mu}(P', s') \right]^* \left[  
       e(P, s) \cdot q \right] - e^{\mu}(P, s) \left[ e^*(P', s') \cdot q 
       \right] \right\} \nonumber \\
       & + & e_{\alpha}^*(P', s') ~ B_{(1)}^{\mu, \alpha \beta} ~ 
       e_{\beta}(P, s) ~ .
       \label{eq:FF1V}
 \ee
For the angular condition one has \cite{Karmanov_rho}
 \be
       (1 + 2 \eta) J_{(1)}^+(1, 1) + J_{(1)}^+(1, -1) & - & \sqrt{8 \eta} 
       J_{(1)}^+(1, 0) - J_{(1)}^+(0, 0) = \nonumber \\
       & = & - B_5^{(1)}(Q^2) - B_7^{(1)}(Q^2) \neq 0 ~ ,
       \label{eq:angular_B}
 \ee
which implies that at $q^+ = 0$ the loss of rotational covariance of the one-body current (\ref{eq:one-body}), i.e. the violation of the angular condition (\ref{eq:angular}), is described by the spurious structures containing $B_5^{(1)}(Q^2)$ and $B_7^{(1)}(Q^2)$ in Eq. (\ref{eq:BV_LF}). Finally, note that the loss of gauge invariance of the one-body current (\ref{eq:one-body}) at $\omega \cdot q = q^+ = 0$ is described by  the spurious form factor $B_8^{(1)}(Q^2)$ appearing in Eq. (\ref{eq:BV_LF}).

\indent As explained in Ref. \cite{DS}, in the Breit frame where $q^+ = 0$ the physical one-body form factors $F_i^{(1)}(Q^2)$ ($i = 1, 2, 3$) can be obtained through the following equations
 \be
       F_1^{(1)}(Q^2) & = & {T_{(1)}^{+, yy} \over 2P^+} ~, \nonumber \\
       F_2^{(1)}(Q^2) & = & {1 \over 2\eta} {T_{(1)}^{+, yy} - T_{(1)}^{+, 
       xx} \over 2P^+} + {1 \over 2 (1 + \eta)} {T_{(1)}^{+, ++} \over 2P^+} 
       - {1 \over \sqrt{\eta (1 + \eta)}} {T_{(1)}^{+, x+} \over 2P^+} ~, 
       \nonumber \\
       F_3^{(1)}(Q^2) & = & {T_{(1)}^{y, xy} \over Q} - \sqrt{{1 + \eta 
       \over \eta}} {T_{(1)}^{y, +y} \over Q} ~.
       \label{eq:solution}
 \ee
At variance with the $PS$ case the components of the tensor (\ref{eq:T1}) with $\mu = y$ are essential for the extraction of the physical form factors, more precisely for the determination of $F_3^{(1)}(Q^2)$. Note also that the form factors $F_1^{(1)}(Q^2)$ and $F_2^{(1)}(Q^2)$ are determined only by the  components with $\mu = +$.

\indent We stress that by means of Eq. (\ref{eq:solution}) the extraction of the physical form factors $F_i^{(1)}(Q^2)$ is not plagued at all by spurious effects, including those related to the loss of rotational covariance and gauge invariance. The angular condition problem (\ref{eq:angular_B}) is therefore completely overcome without introducing explicitly any covariant current operator.

\indent As far the approach of Ref. \cite{LPS} is concerned, the matrix elements $j^{\mu}(s', s)$ of the covariant current (\ref{eq:LPS}) can be expressed in terms of the $LF$ wave function (\ref{eq:wfV}-\ref{eq:Melosh_V}) as: $j^{\mu}(s', s) =$ $\langle P', s' | j^{\mu} | P, s \rangle$ with $| P, s \rangle = e_{\beta}(P, s) ~ | P, \beta \rangle_{LF}$. In Ref. \cite{LPS} it is {\em assumed} that the matrix elements $j^{\mu}(s', s)$ have the following decomposition
\be
       j^{\mu}(s', s) & = & - (P + P')^{\mu} \left\{ {\cal{F}}_1^{(1)}
       e^*(P', s')  \cdot e(P, s) + {{\cal{F}}_2^{(1)} \over 2M_V^2} \left[ 
       e^*(P', s') \cdot q \right] \left[ e(P, s) \cdot q \right] \right\} 
       \nonumber \\
       & + & {\cal{F}}_3^{(1)} \left\{ \left[ e^{\mu}(P', s') \right]^* 
       \left[  e(P, s) \cdot q \right] - e^{\mu}(P, s) \left[ e^*(P', s') 
       \cdot q \right]  \right\}
       \label{eq:FFV_jmu}
 \ee
and it  is shown that there are only three independent matrix elements $j^{\mu}(s', s)$, namely $j_{0 0}^+$, $j_{1 1}^+$ and $j_{1 0}^x$\footnote{Note that the matrix element $j_{1 0}^y$ is not independent from $j_{1 0}^x$ because one has $j_{1 0}^y = i ~ j_{1 0}^x$ \cite{LPS}.}. Thanks to the choice (\ref{eq:LPS_J}), the latter are very simply related to the matrix elements of the one-body current (\ref{eq:one-body}) evaluated at $q^+ \neq 0$, which will be denoted by ${\cal{J}}_{(1)}^{\mu}(s', s)$ to distinguish them from the matrix elements  $J_{(1)}^{\mu}(s', s)$ evaluated at $q^+ = 0$. Ignoring for the moment the possible presence of spurious structures, in the Breit frame where $q^+ = - q^- = Q$ one has
 \be
       j_{0 0}^+ & = & {\cal{J}}_{(1)}^+(0 0) = 2 M_V ~ \sqrt{1 + \eta} 
       \left\{ (1 + 2 \eta) ~ {\cal{F}}_1^{(1)} - 2 \eta ~ (1 + \eta)  ~ 
       {\cal{F}}_2^{(1)} - 2 \eta ~ {\cal{F}}_3^{(1)} \right\} ~ , 
       \nonumber \\
       j_{1 1}^+ & = & {\cal{J}}_{(1)}^+(1 1) = 2 M_V ~ \sqrt{1 + \eta} ~ 
       {\cal{F}}_1^{(1)} ~ , \nonumber \\
       j_{1 0}^x & = & {1 \over 2} \left[ {\cal{J}}_{(1)}^x(1 0) - 
       {\cal{J}}_{(1)}^x(0 1) \right] = 2 M_V ~ \sqrt{1 + \eta} ~ \sqrt{\eta 
       / 2} ~ {\cal{F}}_3^{(1)} ~ .
      \label{eq:LPS_V}
 \ee
Thus, the form factors ${\cal{F}}_i^{(1)}$ ($i = 1, 2, 3$) may be uniquely determined through the following relations
 \be
       {\cal{F}}_1^{(1)}(Q^2 / M_V^2) & = & {1 \over 2M_V \sqrt{1 + \eta}} ~ 
       {\cal{J}}_{(1)}^+(1 1) ~ , \nonumber \\
       {\cal{F}}_2^{(1)}(Q^2 / M_V^2) & = & {1 \over 4 M_V \eta ~ (1 + 
       \eta)} ~ \left\{ {\cal{J}}_{(1)}^+(0 0) - (1 + 2 \eta) ~ 
       {\cal{J}}_{(1)}^+(1 1) \right. \nonumber \\
       & + & \left. \sqrt{2 \eta} ~ \left[ {\cal{J}}_{(1)}^x(1 0) - 
       {\cal{J}}_{(1)}^x(0 1) \right] \right\} ~ , \nonumber \\
       {\cal{F}}_3^{(1)}(Q^2 / M_V^2) & = & {1 \over 2M_V \sqrt{1 + \eta} 
       \sqrt{2 \eta}} ~  \left[ {\cal{J}}_{(1)}^x(1 0) - {\cal{J}}_{(1)}^x(0 
       1) \right]  ~ ,
       \label{eq:FFV_LPS}
 \ee
where we have explicitly taken into account that the form factors ${\cal{F}}_i^{(1)}$ are not function of $Q^2$, but of the ratio $Q^2 / M_V^2$, thanks also to the fact that the matrix elements ${\cal{J}}_{(1)}^{\mu}(s', s)$ are proportional to $2P^+$ because of the normalization of the $LF$ center-of-mass states. Note that the form factor ${\cal{F}}_3^{(1)}$ is determined through the $x$ component of the one-body current, while  ${\cal{F}}_1^{(1)}$ depends on the {\em plus} component only.

\indent The predictions of the $LF$ approaches at $q^+ = 0$ [Eq. (\ref{eq:solution})] and at $q^+ \neq 0$ [Eq. (\ref{eq:FFV_LPS})], obtained in case of the $\rho$ meson adopting for $w_{\rho}(k)$ the corresponding eigenfunction of the quark potential model of Ref. \cite{GI}, are compared in Figs. 4-6 in terms of the conventional charge $G_0(Q^2)$, magnetic $G_1(Q^2)$ and quadrupole $G_2(Q^2)$ form factors, defined as
 \be
       G_0(Q^2) & = & F_1(Q^2) + {2 \eta \over 3} \left[ F_1(Q^2) - F_3(Q^2) 
       - (1 + \eta) F_2(Q^2) \right] ~ , \nonumber \\
       G_1(Q^2) & = & F_3(Q^2) ~ , \nonumber \\
       G_2(Q^2) & = & {\sqrt{8} \eta \over 3} \left[ F_1(Q^2) - F_3(Q^2) 
       - (1 + \eta) F_2(Q^2) \right] ~.
       \label{eq:Gi}
 \ee

\begin{figure}[htb]

\vspace{0.25cm}

\centerline{\epsfxsize=12cm \epsfig{file=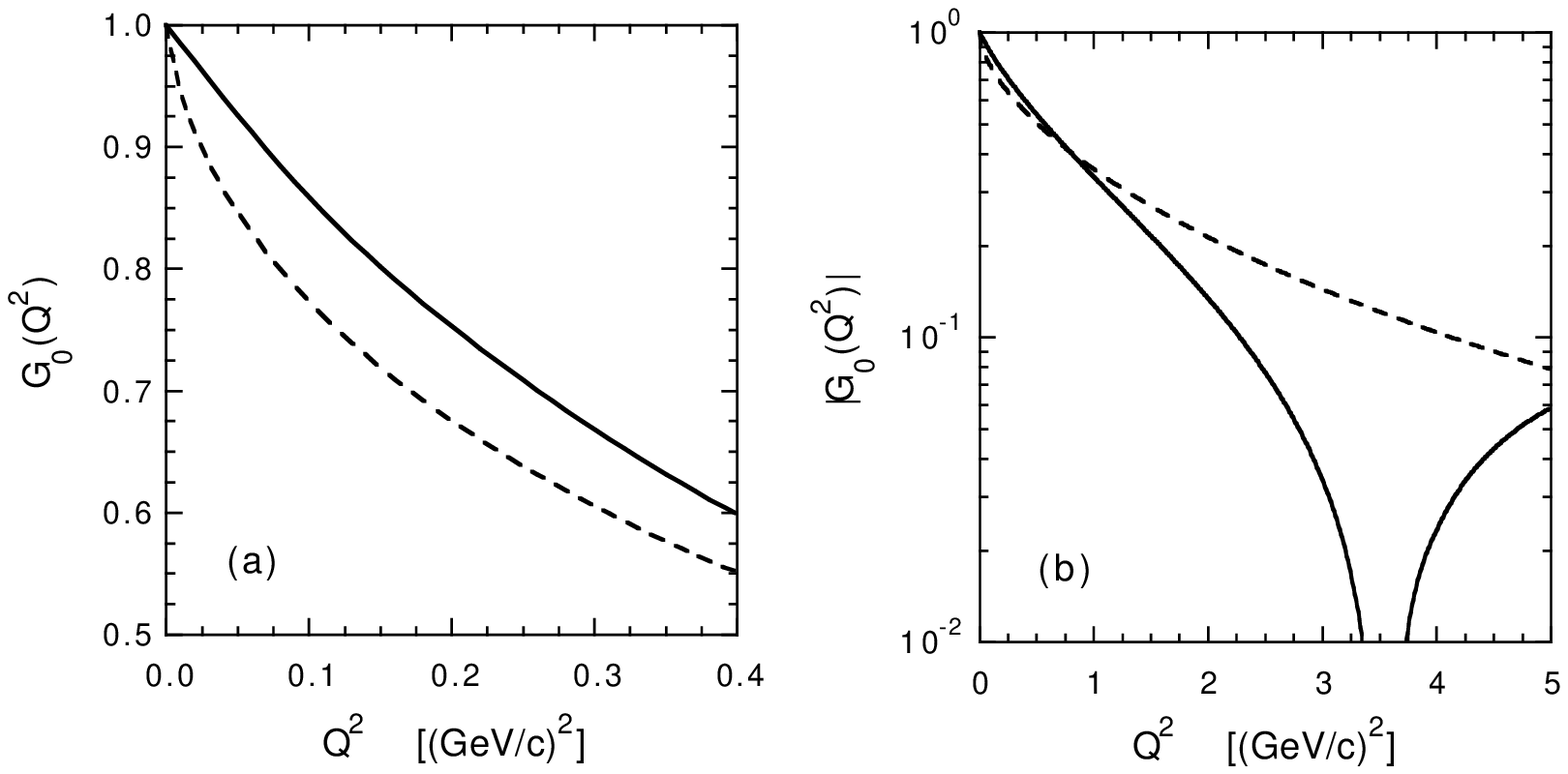}}

{\small \noindent {\bf Figure 4}. Charge form factor $G_0(Q^2)$ for low (a) and high (b) values of $Q^2$. The solid line is the result obtained within the $LF$ approach at $q^+ = 0$  [Eq. (\ref{eq:solution})], while the dashed line is the prediction of the $LF$ approach at $q^+ \neq 0$ [Eq. (\ref{eq:FFV_LPS})]. Point-like constituent quarks are assumed in the calculations. The radial wave function $w_{\rho}(k)$ is taken to be the corresponding eigenfunction of the quark potential model of Ref. \cite{GI}.}

\vspace{0.25cm}

\end{figure}

\begin{figure}[htb]

\vspace{0.25cm}

\centerline{\epsfxsize=12cm \epsfig{file=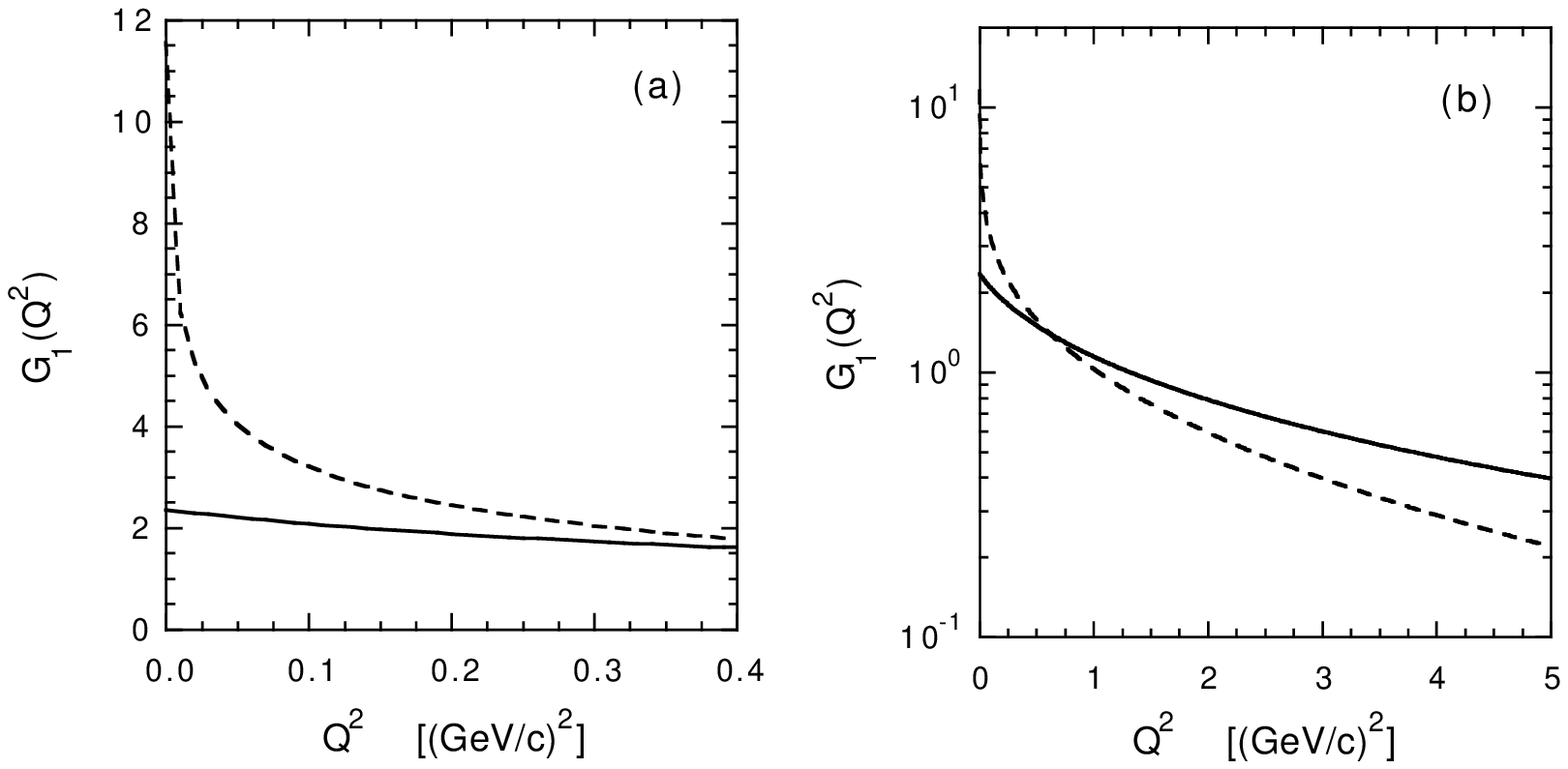}}

{\small \noindent {\bf Figure 5}. The same as in Fig. 4, but for the magnetic form factor $G_1(Q^2)$.}

\vspace{0.25cm}

\end{figure}

\begin{figure}[htb]

\vspace{0.25cm}

\centerline{\epsfxsize=12cm \epsfig{file=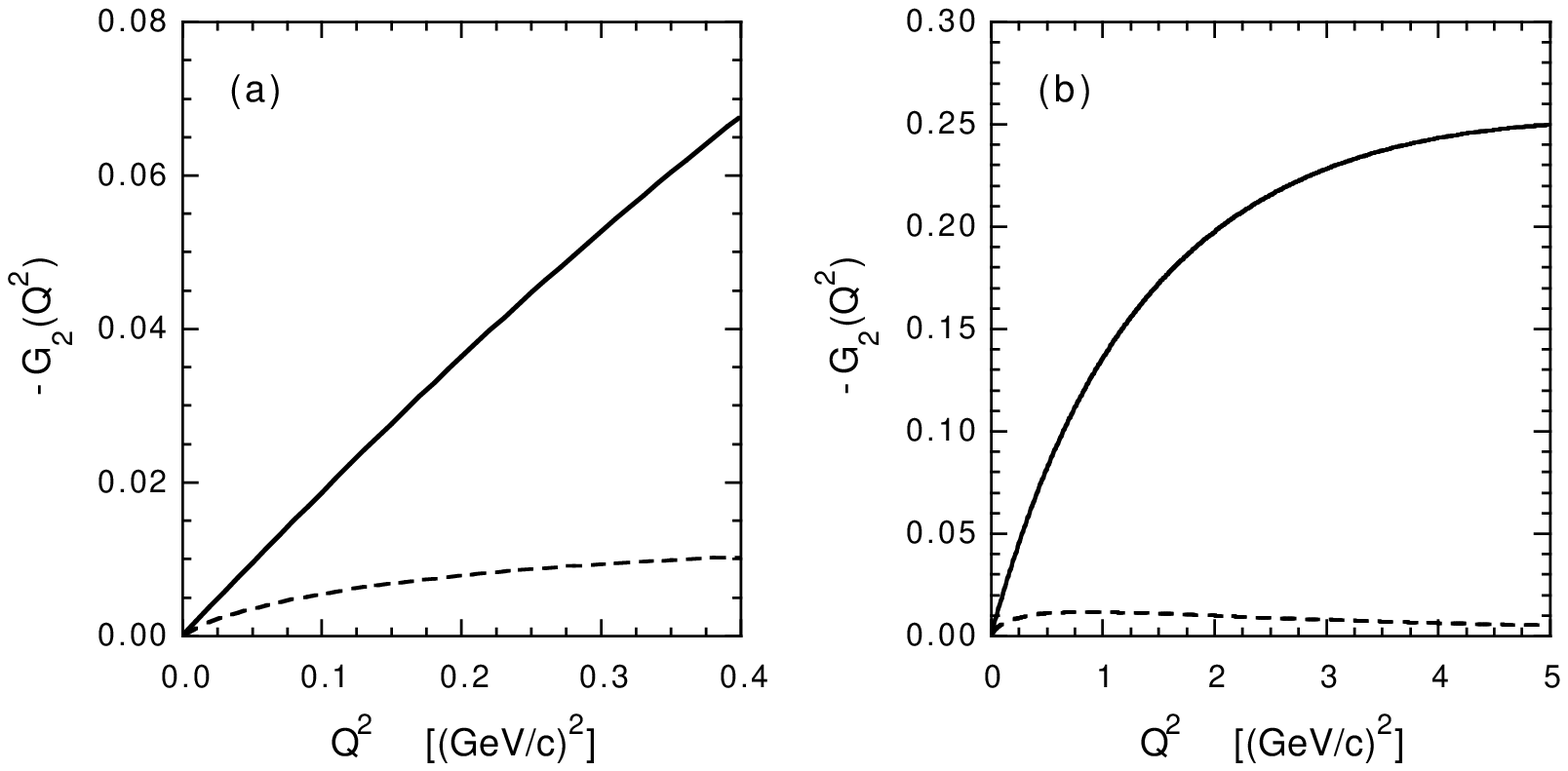}}

{\small \noindent {\bf Figure 6}. The same as in Fig. 4, but for the quadrupole form factor $-G_2(Q^2)$.}

\vspace{0.25cm}

\end{figure}

\indent It can be seen that: ~ i) the charge radius, $r_{ch} \equiv \sqrt{- 6 dG_0(Q^2) / dQ^2 |_{Q^2 = 0}}$, turns out to be $r_{ch} = 1.1 ~ fm$ at $q^+ \neq 0$ and $r_{ch} = 0.57 ~ fm$ at $q^+ = 0$\footnote{Taking into account a constituent size of $\simeq 0.45 ~ fm$ as in the pion and kaon cases, the $LF$ approach at $q^+ = 0$ predicts a charge radius of the $\rho$ meson equal to $\simeq 0.75 ~ fm$.}; ~ ii) the charge form factor calculated at $q^+ = 0$ exhibits a node at $Q^2 \simeq 3.4 ~ (GeV/c)^2$ at variance with the result obtained at $q^+ \neq 0$; ~ iii) the magnetic moment, $\mu_{\rho} \equiv G_1(Q^2 = 0)$, is remarkably larger at $q^+ \neq 0$ [$\mu_{\rho} = 11.6$] than the corresponding one at $q^+ = 0$ [$\mu_{\rho} = 2.35$], which is quite close to the non-relativistic limit $\mu_{\rho} = 2$ (cf. Ref. \cite{DS}); moreover, the magnetic form factor $G_1(Q^2)$ calculated at $q^+ \neq 0$ has a sharp upturn near the photon point completely at variance with the result obtained at $q^+ = 0$; ~ iv) the quadrupole form factor $G_2(Q^2)$ is predicted to be quite small at $q^+ \neq 0$ both at low and high values of $Q^2$. All these findings are a direct manifestation of the lack of the $Z$-graph in the $LF$ approach at $q^+ \neq 0$. Thus, for the $\rho$ meson the $Z$-graph plays an important role both at low and high values of $Q^2$. We want to point out that the node obtained in the charge form factor $G_0(Q^2)$ at $q^+ = 0$ may be (at least partially) related to the nature of the spin-spin term of the quark potential model of Ref. \cite{GI}, which is repulsive in the triplet spin state at short interquark distances.

\indent Let us now consider the heavy quark limit in which only one quark is coupled to the virtual photon and its mass goes to infinity, i.e. $m_1 \to \infty$. Thanks to the $HQS$ we expect that for fixed values of the variable $w \equiv P \cdot P' / M_V^2 = 1 + Q^2 / 2 M_V^2$ one should have
 \be
        \mbox{lim}_{m_1 \to \infty} ~ G_0(Q^2) & = & \xi_{IW}(w) ~ , 
        \nonumber \\
        \mbox{lim}_{m_1 \to \infty} ~ G_1(Q^2) & = & \xi_{IW}(w) ~ , 
        \nonumber \\
        \mbox{lim}_{m_1 \to \infty} ~ G_2(Q^2) & = & 0 ~ ,
        \label{eq:HQlimit_V}
 \ee
where $\xi_{IW}(w)$ is the same $IW$ function encountered in the $PS$ case. We have calculated the form factors $G_i$ ($i = 1, 2, 3$) for various values of the constituent mass $m_1$ pertaining to the cases of $\rho$, $K^*$, $D^*$ and $B^*$ mesons, keeping fixed the constituent mass $m_2$ at the value $m_2 = 0.220 ~ GeV$. The results are reported in Figs. 7-9 in terms of the variable $w$ and compared with the $IW$ function $\xi_{IW}(w)$ calculated in Ref. \cite{SIM_IW}.

\begin{figure}[htb]

\vspace{0.25cm}

\centerline{\epsfxsize=12cm \epsfig{file=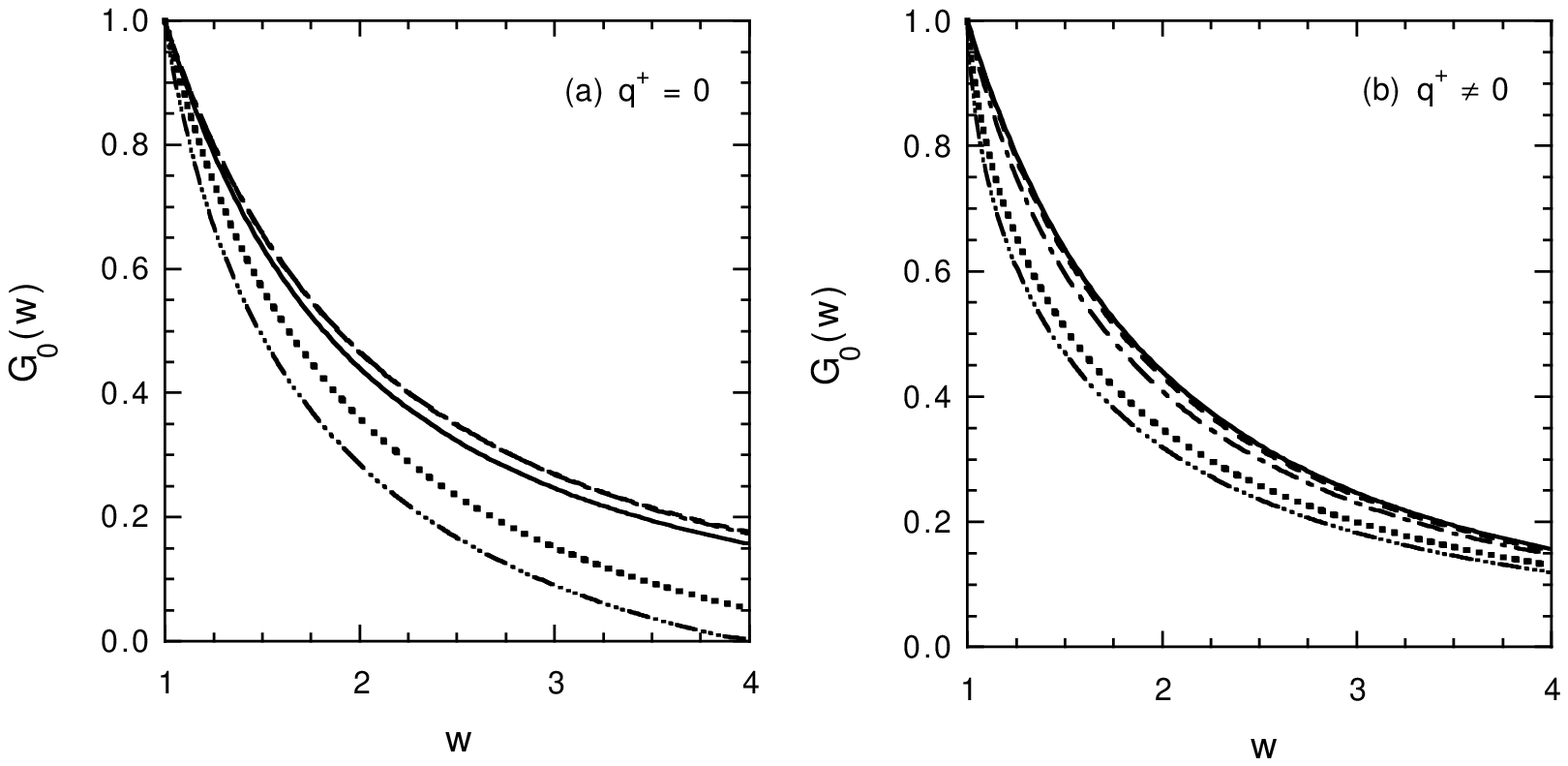}}

{\small \noindent {\bf Figure 7}. Charge form factor $G_0(w)$ versus the variable $w$, calculated within the $LF$ approach of Ref. \protect\cite{DS} at $q^+ = 0$ (a), corresponding to Eq. (\protect\ref{eq:solution}), and the one of Ref. \protect\cite{LPS} at $q^+ \neq 0$ (b), corresponding to Eq. (\protect\ref{eq:FFV_LPS}) . Triple-dot-dashed, dotted, dot-dashed and dashed lines correspond to the cases of $\rho$, $K^*$, $D^*$ and $B^*$ mesons, respectively. In (a) and (b) the solid line is the same $IW$ function $\xi_{IW}(w)$ as calculated in Ref. \protect\cite{SIM_IW}. In (a) the dot-dashed and dashed lines almost coincide. Point-like constituent quarks are assumed in the calculations. The radial wave functions $w_V(k)$ are taken to be the eigenfunctions of the quark potential model of Ref. \protect\cite{GI}.}

\vspace{0.25cm}

\end{figure}

\begin{figure}[htb]

\vspace{0.25cm}

\centerline{\epsfxsize=12cm \epsfig{file=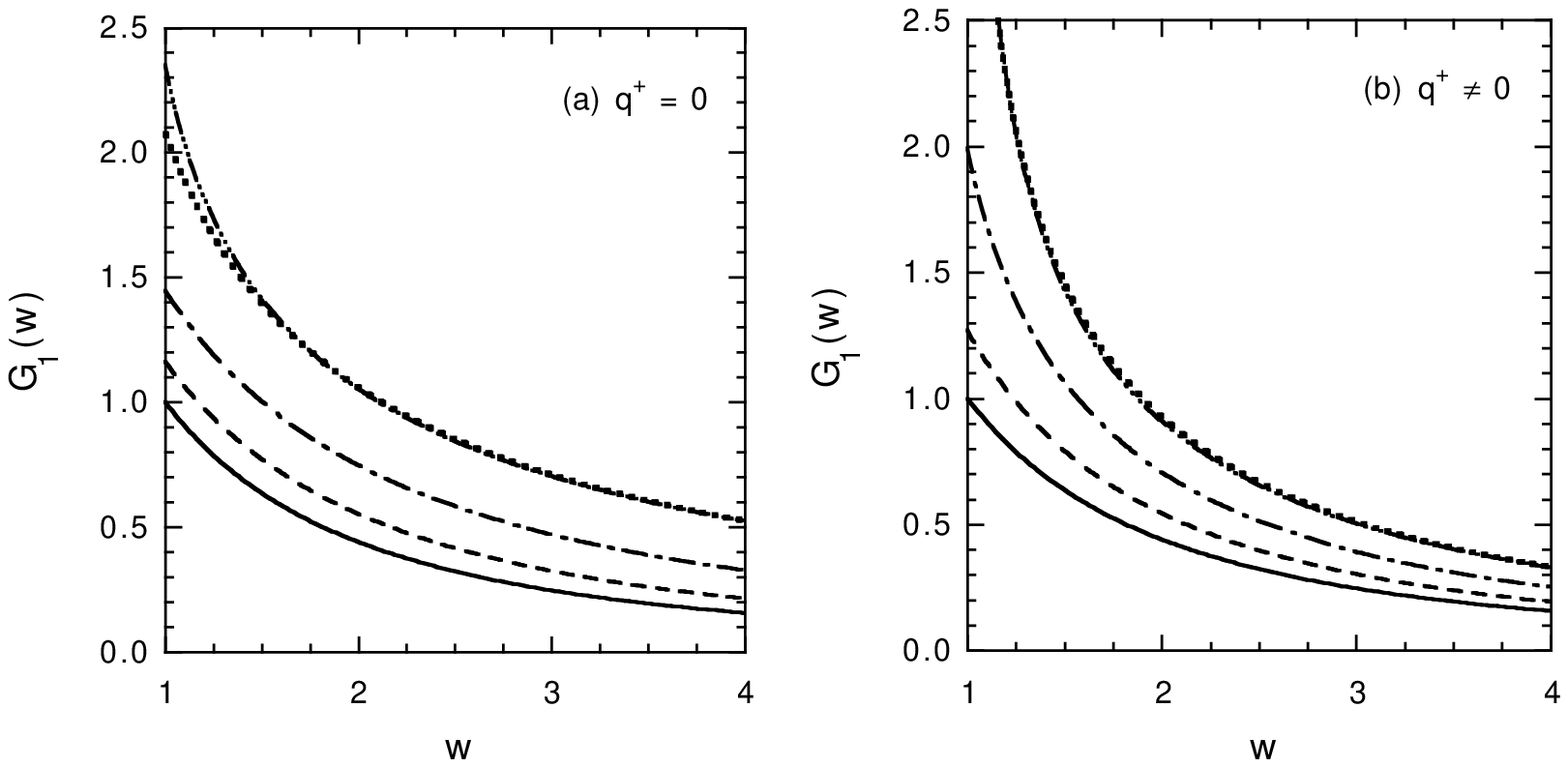}}

{\small \noindent {\bf Figure 8}. The same as in Fig. 7, but for the magnetic form factor $G_1(w)$.}

\vspace{0.25cm}

\end{figure}

\begin{figure}[htb]

\vspace{0.25cm}

\centerline{\epsfxsize=12cm \epsfig{file=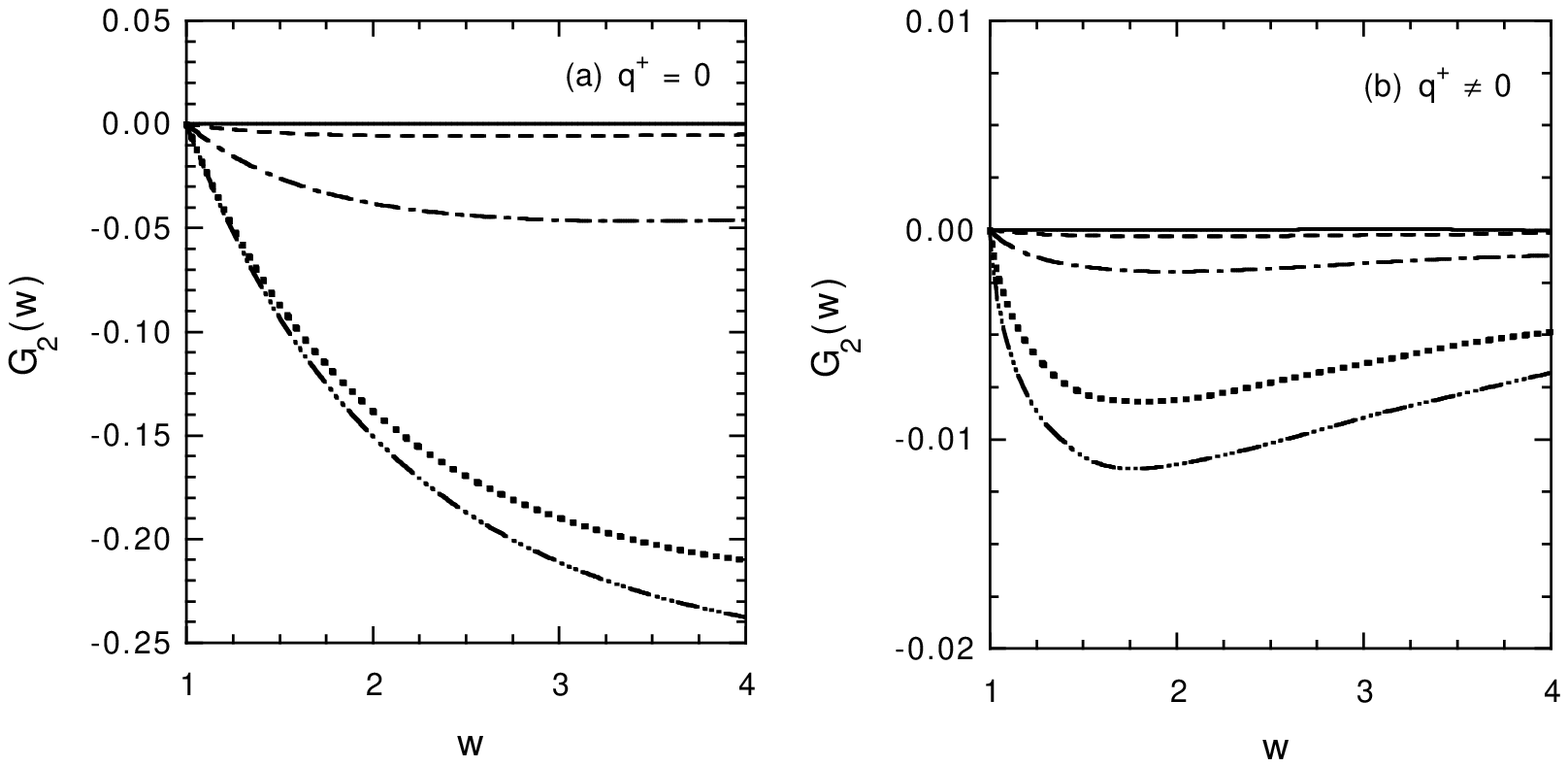}}

{\small \noindent {\bf Figure 9}. The same as in Fig. 7, but for the quadrupole form factor $G_2(w)$.}

\vspace{0.25cm}

\end{figure}

\indent It can clearly be seen that Eq. (\ref{eq:HQlimit_V}) is fulfilled, confirming in this way that also in case of vector mesons the differences between the results at $q^+ = 0$ and $q^+ \neq 0$ are due to the lack of the $Z$-graph contribution at $q^+ \neq 0$. We want to stress that the universality of the calculated $IW$ function, implied by the fulfillment of Eqs. (\ref{eq:HQlimit_PS}) and (\ref{eq:HQlimit_V}), is in nice agreement with the predictions of the $HQS$, which is an exact symmetry of $QCD$ in the heavy-quark limit. Finally, from Figs. 8-9 a couple of comments are in order: ~ i) the difference in the calculated magnetic moment is still remarkably sizable at the $D^*$ mass (i.e., around $M_V \simeq 2 ~ GeV$), and ~ ii) the lack of the $Z$-graph at $q^+ \neq 0$ makes the quadrupole form factor $G_2$ very small even at quite large values of the (active) quark mass.

\indent Let us now consider the question whether the decomposition (\ref{eq:FFV_jmu}) is complete without introducing spurious structures analogous to those appearing at $q^+ = 0$ in Eqs. (\ref{eq:BV_LF}-\ref{eq:FF1V}). It is clear that the dependence of $e.m.$ amplitudes upon the four-vector $\omega$, which we remind identifies the orientation of the null hyperplane where the $LF$ wave function is defined, can occur independently of the value of $q^+$. Therefore, at $q^+ \neq 0$ one expects that the following general decomposition holds for the matrix elements ${\cal{J}}_{(1)}(s', s)$ of the one-body current:
\be
       {\cal{J}}_{(1)}(s', s) & = & - (P + P')^{\mu} \left\{ 
       {\cal{F}}_1^{(1)} e^*(P', s')  \cdot e(P, s) + {{\cal{F}}_2^{(1)} 
       \over 2M_V^2} \left[ e^*(P', s') \cdot q \right] \left[ e(P, s) \cdot 
       q \right] \right\} \nonumber \\
       & + & {\cal{F}}_3^{(1)} \left\{ \left[ e^{\mu}(P', s') \right]^* 
       \left[  e(P, s) \cdot q \right] - e^{\mu}(P, s) \left[ e^*(P', s') 
       \cdot q \right]  \right\} \nonumber \\
       & + & e_{\alpha}^*(P', s') ~ {\cal{B}}_{(1)}^{\mu, \alpha \beta} ~ 
       e_{\beta}(P, s) 
       \label{eq:J1}
 \ee
with
 \be
       {\cal{B}}_{(1)}^{\mu, \alpha \beta}(\omega) & = & {M_V^2 \over 2 
       (\omega \cdot \tilde{P})} \omega^{\mu} ~ \left[ {\cal{B}}_1^{(1)} ~ 
       g^{\alpha \beta} + {\cal{B}}_2^{(1)} ~ {q^{\alpha} q^{\beta} \over 
       M_V^2} + M_V^2 {\cal{B}}_3^{(1)} ~ {\omega^{\alpha} \omega^{\beta} 
       \over (\omega \cdot \tilde{P})^2} \right. \nonumber \\
       & + & \left. {\cal{B}}_4^{(1)} ~ {q^{\alpha} \omega^{\beta} - 
       q^{\beta} \omega^{\alpha} \over 2 (\omega \cdot \tilde{P}}) \right] + 
       (P + P')^{\mu} \left[ M_V^2 {\cal{B}}_5^{(1)} ~ {\omega^{\alpha} 
       \omega^{\beta} \over (\omega \cdot \tilde{P})^2} \right.
       \nonumber \\
       & + & \left. {\cal{B}}_6^{(1)} ~ {q^{\alpha} \omega^{\beta} - 
       q^{\beta} \omega^{\alpha} \over 2 (\omega \cdot \tilde{P})} \right] + 
       {M_V^2 \over (\omega \cdot \tilde{P})} ~ {\cal{B}}_7^{(1)} ~ \left[ 
       \left( g^{\mu \alpha} - {q^{\mu} q^{\alpha} \over q^2} \right) 
       \omega^{\beta} \right. \nonumber \\
       & + & \left. \left( g^{\mu \beta} - {q^{\mu} q^{\beta} \over q^2} 
       \right) \omega^{\alpha} \right] + {\cal{B}}_8^{(1)} ~ q^{\mu} ~ 
       {q^{\alpha} \omega^{\beta} + q^{\beta} \omega^{\alpha} \over 2 
       (\omega \cdot \tilde{P})} ~ ,
       \label{eq:B1}
 \ee
where $\tilde{P} \equiv (P + P')/2$. With respect to Eq. (\ref{eq:BV_LF}) we have introduced a different notation for the spurious form factors, because the latter may depend in general on $\omega \cdot q$, i.e. they can be different at $q^+ = 0$ and $q^+ \neq 0$. Note that at $q^+ \neq 0$ the loss of gauge invariance of the one-body current is described not only by the spurious structure containing ${\cal{B}}_8^{(1)}$ (as in the case $q^+ = 0$), but also by the ones proportional to ${\cal{B}}_k^{(1)}$ with $k = 1, ..., 4$.

\indent By means of the four-vector $\omega$ the choice (\ref{eq:LPS_J}) for the operator $C^{\mu}$ can be conveniently cast into the following form: $C^{\mu} = J_{(1)}^{\mu} - \omega^{\mu} (q \cdot J_{(1)}) / (\omega \cdot q)$. Therefore, using Eqs. (\ref{eq:J1}-\ref{eq:B1}) one immediately obtains that the spurious structures containing the form factors ${\cal{B}}_k^{(1)}$ with $k = 1, ..., 4$ cannot appear in the decomposition of the matrix elements $C^{\mu}(s', s)$ of the gauge-invariant operator $C^{\mu}$. Consequently, also for the matrix elements $j^{\mu}(s', s)$ we generally expect that 
\be
       j^{\mu}(s', s) & = & - (P + P')^{\mu} \left\{ {\cal{F}}_1^{(1)}
       e^*(P', s')  \cdot e(P, s) + {{\cal{F}}_2^{(1)} \over 2M_V^2} \left[ 
       e^*(P', s') \cdot q \right] \left[ e(P, s) \cdot q \right] \right\} 
       \nonumber \\
       & + & {\cal{F}}_3^{(1)} \left\{ \left[ e^{\mu}(P', s') \right]^* 
       \left[  e(P, s) \cdot q \right] - e^{\mu}(P, s) \left[ e^*(P', s') 
       \cdot q \right]  \right\} \nonumber \\
       & + & e_{\alpha}^*(P', s') ~ \overline{\cal{B}}_{(1)}^{\mu, \alpha 
       \beta} ~ e_{\beta}(P, s) 
       \label{eq:jmu}
 \ee
with
 \be
       \overline{\cal{B}}_{(1)}^{\mu, \alpha \beta}(\omega) & = & 
       (P + P')^{\mu} \left[ M_V^2 {\cal{B}}_5^{(1)} ~ {\omega^{\alpha} 
       \omega^{\beta} \over (\omega \cdot \tilde{P})^2} + {\cal{B}}_6^{(1)} 
       ~ {q^{\alpha} \omega^{\beta} - q^{\beta} \omega^{\alpha} \over 2 
       (\omega \cdot \tilde{P})} \right] \nonumber \\
       & + & {M_V^2 \over (\omega \cdot \tilde{P})} ~ {\cal{B}}_7^{(1)} ~ 
       \left[ \left( g^{\mu \alpha} - {q^{\mu} q^{\alpha} \over q^2} \right) 
       \omega^{\beta} + \left( g^{\mu \beta} - {q^{\mu} q^{\beta} \over q^2} 
       \right) \omega^{\alpha} \right] \nonumber \\
       & + & {\cal{B}}_8^{(1)} ~ \left[ q^{\mu} - \omega^{\mu} {q^2 \over 
       (\omega \cdot q)} \right] ~ {q^{\alpha} \omega^{\beta} + q^{\beta} 
       \omega^{\alpha} \over 2 (\omega \cdot \tilde{P})} ~ .
       \label{eq:B1_jmu}
 \ee

\indent In order to prove that the decomposition (\ref{eq:FFV_jmu}) is not complete, it is enough to demonstrate that at least one of the four spurious form factors ${\cal{B}}_{5-8}^{(1)}$ is non vanishing. To this end let us first consider that the possible presence of spurious structures corresponds to replace Eq. (\ref{eq:LPS_V}) with
\be
       j_{0 0}^+ & = & {\cal{J}}_{(1)}^+(0 0) = 2 M_V ~ \sqrt{1 + \eta} 
       \left\{ (1 + 2 \eta) ~ {\cal{F}}_1^{(1)} - 2 \eta ~ (1 + \eta)  ~ 
       {\cal{F}}_2^{(1)} - 2 \eta ~ {\cal{F}}_3^{(1)} \right. \nonumber \\
       & + & \left. {{\cal{B}}_5^{(1)} \over 1 + \eta} + 2 \eta 
       {\cal{B}}_6^{(1)} - {\eta \over 1 + \eta} {\cal{B}}_7^{(1)} - 2 \eta 
       {\cal{B}}_8^{(1)} \right\} ~ , 
       \nonumber \\
       j_{1 1}^+ & = & {\cal{J}}_{(1)}^+(1 1) = 2 M_V ~ \sqrt{1 + \eta} ~ 
       {\cal{F}}_1^{(1)} ~ , \nonumber \\
       j_{1 0}^x & = & {1 \over 2} \left[ {\cal{J}}_{(1)}^x(1 0) - 
       {\cal{J}}_{(1)}^x(0 1) \right] = 2 M_V ~ \sqrt{1 + \eta} ~ \sqrt{\eta 
       / 2} ~ \left\{ {\cal{F}}_3^{(1)} + {1 \over 2 (1 + \eta)} 
       {\cal{B}}_7^{(1)} \right\} ~ .
      \label{eq:LPS_B}
 \ee
Therefore, the extraction of the form factors ${\cal{F}}_2^{(1)}$ and ${\cal{F}}_3^{(1)}$ may be plagued by spurious effects, while only ${\cal{F}}_1^{(1)}$ is free from spurious effects. We want to point out that the relation $j_{1 0}^y = i ~ j_{1 0}^x$ is not modified by the possible presence of spurious structures in Eqs. (\ref{eq:jmu}-\ref{eq:B1_jmu}); in other words the number of independent matrix elements of the current operator $j^{\mu}$ is still three even in presence of the spurious form factors appearing in Eq. (\ref{eq:B1_jmu}).

\begin{figure}[htb]

\vspace{0.25cm}

\centerline{\epsfxsize=12cm \epsfig{file=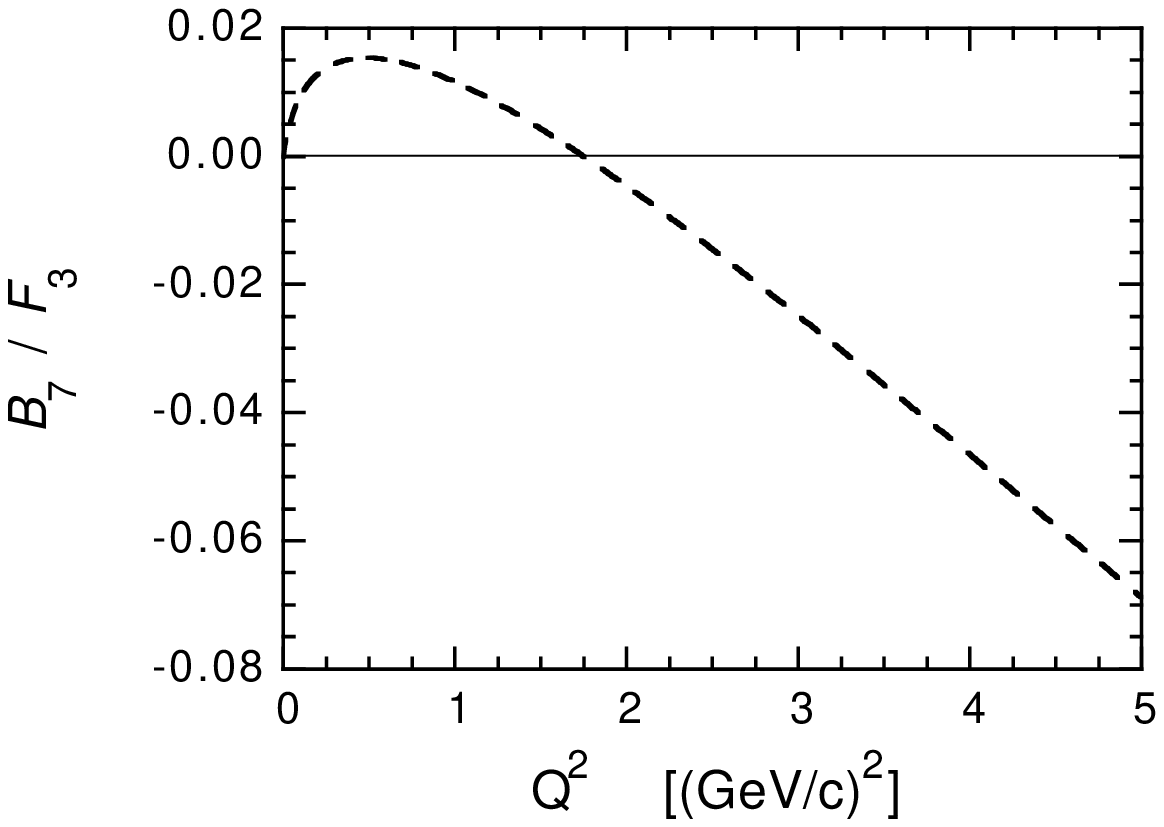}}

{\small \noindent {\bf Figure 10}. Ratio of the spurious form factor ${\cal{B}}_7^{(1)}$, calculated from Eq. (\ref{eq:B7}), to the magnetic form factor ${\cal{F}}_3^{(1)}$, obtained from Eq. (\ref{eq:FFV_LPS}) in case of the $\rho$ meson within the approach of Ref. \cite{LPS} at $q^+ \neq 0$ (dashed line).}

\vspace{0.25cm}

\end{figure}

\indent In addition to Eq. (\ref{eq:LPS_B}) one has
 \be
        {1 \over 2} \left[ {\cal{J}}_{(1)}^x(1 0) +  {\cal{J}}_{(1)}^x(0 1) 
        \right] = - {M_V \over \sqrt{2}} {\cal{B}}_7^{(1)} ~ , 
        \label{eq:B7}
 \ee
which allows us to calculate the spurious form factor ${\cal{B}}_7^{(1)}$ directly in terms of matrix elements of the one-body current. The explicit calculation of the {\em l.h.s.} of Eq. (\ref{eq:B7}), carried out for the case of the $\rho$ meson, is reported in Fig. 10. It can be seen that ${\cal{B}}_7^{(1)}$ is just a small fraction of ${\cal{F}}_3^{(1)}$ and therefore its impact on the extraction of ${\cal{F}}_3^{(1)}$ is quite limited. However, what really matters is not the quantitative impact of ${\cal{B}}_7^{(1)}$, but the conceptual fact that it is non vanishing, which demonstrates that the covariant decomposition (\ref{eq:FFV_jmu}) is not complete. Note that at the photon point ${\cal{B}}_7^{(1)} = 0$, so that the anomalous result $\mu_{\rho} = 11.6$ obtained at $q^+ \neq 0$ does not depend on spurious effects. Furthermore, the other spurious form factors  ${\cal{B}}_5^{(1)}$, ${\cal{B}}_6^{(1)}$ and ${\cal{B}}_8^{(1)}$ entering Eq. (\ref{eq:LPS_B}), cannot be calculated in terms of matrix elements of the one-body current. Thus we can conclude that the approach of Ref. \cite{LPS} not only ignores the spurious terms arising from the orientation of the null hyperplane, but also it is not able to eliminate consistently such spurious effects in the extraction of the physical form factors.

\section{Conclusions}

\indent In this paper we have carried out  a detailed comparison of the predictions of the light-front approaches of Refs. \cite{DS} and \cite{LPS} in case of the (space-like) electromagnetic elastic form factors of both light and heavy pseudoscalar and vector mesons adopting the general framework of the constituent quark model. The two approaches are based on the one-body approximation (\ref{eq:one-body}) for the electromagnetic current operator, but they are elaborated in different Breit frames, namely at $q^+ = 0$ \cite{DS} and $q^+ \neq 0$ \cite{LPS}. It has been shown that: ~ i) the two light-front approaches are inequivalent because of the different contribution of the $Z$-graph at $q^+ = 0$ or $q^+ \neq 0$. While at $q^+ = 0$ it is possible to cancel out exactly the $Z$-graph (see Ref. \cite{DS}), the latter is active at $q^+ \neq 0$, but ignored in Ref. \cite{LPS}; ~ ii) the $Z$-graph provides an important contribution in case of light hadrons, whereas it vanishes in the heavy-quark limit, where the two light-front approaches predict the same universal Isgur-Wise function. 

\indent We have also pointed out an important feature of the approach of Ref. \cite{LPS}, namely: the elastic form factors of a hadron are basically function of $Q^2 / M^2$, where $M$ is the mass of the hadron. We have shown that such a dependence (shared also by the point-form approach of Refs. \cite{PFSA,Klink}) is not efficient for describing the phenomenology of light hadrons. In other words, the light-front approach at $q^+ \neq 0$ maximizes the impact of the many-body currents needed to achieve consistency with experiment.

\indent Finally, in case of vector mesons, the spurious effects related to the orientation of the null hyperplane where the $LF$ wave function is defined, have been analyzed in details. While such unwanted effects are properly eliminated in the approach of Ref. \cite{DS}, they are ignored and cannot be eliminated within the approach of Ref. \cite{LPS}. Thus, we can conclude that, as far as the one-body current (\ref{eq:one-body}) is concerned, the predictions of the light-front approach at $q^+ = 0$ should be preferred, particularly in case of light hadrons.

\section*{Acknowledgments} The author gratefully acknowledges E. Pace for many fruitful discussions about the approach of Ref. \cite{LPS}.

\end{document}